\documentclass[acmsmall]{acmart}

\usepackage{booktabs} 

\usepackage{amsmath}
\usepackage{multirow}

\usepackage{multicol}
\usepackage{graphicx}
\usepackage[ruled, linesnumbered]{algorithm2e}
\usepackage[normalem]{ulem}
\usepackage{subcaption}
\setcopyright{rightsretained}
\newcommand{\jrnAdd}[1]{\textcolor{black}{#1}}

\newcommand{\rev}[1]{\textcolor{black}{#1}}
\newcommand{\minrev}[1]{\textcolor{black}{#1}}
\begin{document}
	\sloppy
\title{\rev{An Energy-Aware Online Learning Framework for Resource Management in Heterogeneous Platforms}}

\author{Sumit K. Mandal}
\email{skmandal@asu.edu}
\orcid{0000-0003-1085-2189}
\affiliation{%
	\institution{Arizona State University}
	\department{School of Electrical, Computer, and Energy Engineering}
	\city{Tempe}
	\state{AZ}
	\postcode{85287}
	\country{USA}
}
\author{Ganapati Bhat}
\email{gmbhat@asu.edu}
\affiliation{%
	\institution{Arizona State University}
	\department{School of Electrical, Computer, and Energy Engineering}
	\city{Tempe}
	\state{AZ}
	\postcode{85287}
	\country{USA}
}
\author{Janardhan Rao Doppa}
\email{jana.doppa@wsu.edu}
\affiliation{%
	\institution{Washington State University}
 	\department{School of EECS}
 	\city{Pullman}
	\state{WA}
	\country{USA}
}
\author{Partha Pratim Pande}
\email{pande@wsu.edu}
\affiliation{%
	\institution{Washington State University}
	\department{School of EECS}
 	\city{Pullman}
	\state{WA}
	\country{USA}
}
\author{Umit Y. Ogras}
\email{umit@asu.edu}
\affiliation{%
	\institution{Arizona State University}
	\department{School of Electrical, Computer, and Energy Engineering}
	\city{Tempe}
	\state{AZ}
	\postcode{85287}
	\country{USA}
}

\begin{abstract}

Mobile platforms must satisfy the contradictory requirements of fast response time and minimum energy consumption as a function of dynamically changing applications. 
To address this need, system-on-chips (SoC) that are at the heart of these devices provide a variety of control knobs, such as the number of active cores and their voltage/frequency levels. Controlling these knobs optimally at runtime is challenging for two reasons. 
First, the large configuration space prohibits exhaustive solutions. 
Second, control policies designed offline are at best sub-optimal since many potential  new applications are unknown at design-time. 
We address these challenges by proposing an online imitation learning approach. 
Our key idea is to construct an offline policy and adapt it online to new applications to optimize a given metric (e.g., energy). The proposed methodology leverages the supervision enabled by power-performance models learned at runtime. 
We demonstrate its effectiveness on a commercial mobile platform with 16 diverse benchmarks. Our approach successfully adapts the control policy to an unknown application after executing less than 25\% of its instructions.
\end{abstract}



\thanks{This work was supported partially by USA Army Research Office grant W911NF-17-1-0485, National Science Foundation grants CNS-1526562 and OAC-1910213 and Semiconductor Research Corporation (SRC) task 2721.001.
}

\setcopyright{acmcopyright}
\acmJournal{TODAES}
\acmYear{2020} \acmVolume{1} \acmNumber{1} \acmArticle{1} \acmMonth{1} \acmPrice{15.00}\acmDOI{10.1145/3386359}

\maketitle

\section{Introduction} \label{sec:intro}


Over a billion people use various types of electronic devices including mobile phones, tablets, and personal computers~\cite{Statista2018_apps}. 
As the processing and sensing capabilities of these devices expand, 
we see exponential growth in the number and types of applications. 
Common examples include graphics-intensive games, 
communication-intensive social media apps, 
health monitoring, 
and traditional compute-intensive applications. 
Delivering the required performance on-demand and maximizing the battery life are two common goals independent of the application. 
However, the runtime configurations to achieve these goals 
can vary dramatically for different application scenarios. 
For example, high-performance CPU cores (e.g., big cores) are preferred over low-power cores while running compute-intensive applications. 
Furthermore, the voltage and frequency levels should be controlled optimally at runtime, since the highest (i.e., most power-hungry) levels are not needed continuously. 
Therefore, heterogeneous SoCs must orchestrate the utilization of the available resources optimally at runtime as the composition of active applications evolve. 

Determining the optimal SoC configuration at runtime is challenging for two reasons. 
First, the parameter space is prohibitively large to explore at runtime. 
Even with two types of cores (e.g., $n_B$ big and $n_L$ little cores) with $F$ frequency levels each, there are thousands of possible configurations ($n_B^F \times n_L^F$). 
Second, and more importantly, one cannot find the optimal configuration offline 
since many potential applications are unknown at design-time. 
Even if a set of key applications, i.e., key performance indicators (KPI), are available, we do not know the precise composition of these KPIs and background applications in advance.
Furthermore, the type and number of applications grow continuously. 
Therefore, there is a strong need for approaches that adapt to new applications 
by learning optimal SoC configurations at runtime.

\begin{figure}[t]
	\centering
	\includegraphics[width=0.67\linewidth]{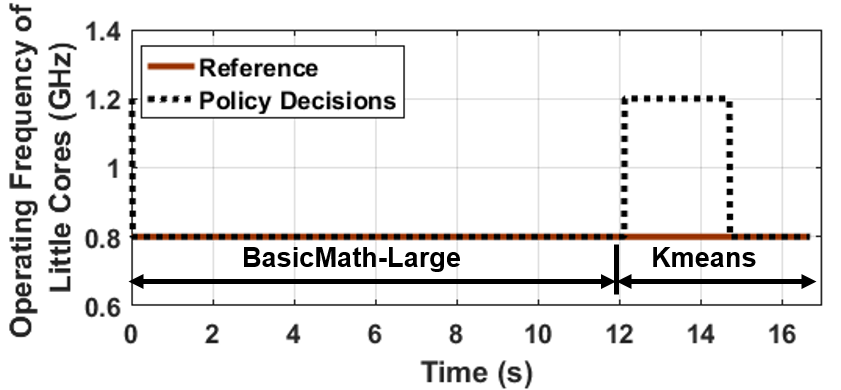}
	\vspace{-3mm}
	\caption{Illustration of the offline policy constructed using a training set that includes BasicMath-Large. The policy decisions for BasicMath-Large are perfect for entire execution,
	while decisions for Kmeans are different 50\% of the time with respect to the reference.}
	\label{fig:policy_pred}
	\vspace{-5mm}
\end{figure}

Existing governors in mobile platforms use simple metrics for runtime power management decisions. For example, interactive and on-demand governors in Android phones control the operating frequency as a function of the utilization. 
These policies maximize performance, but they are not energy-efficient.
Several power management techniques have recently been proposed to overcome these limitations~\cite{martinez2009dynamic, park2017ml, pathania2015power}.
These policies are typically built offline using prior knowledge of few known applications.
Therefore, they may not perform well for new applications encountered at runtime.  
For example, Figure~\ref{fig:policy_pred} illustrates an offline policy which performs poorly while running applications outside the training set.
Reinforcement learning (RL)~\cite{sutton2018reinforcement} methods, such as Q-learning, can be employed for online learning. However, RL methods are not efficient to learn the optimal policy for new applications, 
since they learn via trial-and-error using very weak training signals and require a large number of data samples.


In this paper, we propose a novel online imitation learning ({\em Online-IL}) approach to learn optimal policies for new applications at runtime. The Online-IL approach leverages an offline control algorithm, constructed at design-time, as the initial policy to effectively bootstrap the learning process.
This policy may not perform well on new applications seen at runtime as demonstrated by  Figure~\ref{fig:policy_pred} and our experimental results. 
Hence, the proposed Online-IL approach synergistically combines the benefits of power/performance models and machine learning techniques. 
The power/performance models provide strong supervision via low-overhead and accurate evaluation of candidate configurations.
Our machine learning techniques leverage this supervision to efficiently adapt the policy to new applications. In each training epoch, we perform the following four steps.
1) Execute the current policy and record the workload metadata, such as the number of instructions and memory accesses; 
2) Evaluate the candidate configurations using the metadata and analytical power/performance models to determine the best action; 
3) Update the policy parameters such that evaluation of policy at each system state matches the best configuration found via power/performance models; and 
4) If required, update the parameters of analytical power/performance models based on real measurements. We repeat these four steps until the policy convergences.

Experimental evaluations on Odroid-XU3 board show that offline policies achieve almost 100\% accuracy with respect to a golden reference while running applications from the training set. 
However, they perform poorly under new applications seen at runtime.
Our proposed online-IL methodology adapts the offline policy online to new applications within a few seconds.

\vspace{1mm}
\noindent {\bf Contributions:} The main contributions of this work are:
\begin{itemize}
	\item A novel approach to utilize online power/performance models at runtime to generate training data,
	
	\item A runtime adaptive control algorithm that combines optimal offline policies and online learning,
	
	\item Comprehensive experimental evaluation on a commercial mobile platform using 16 diverse applications.
\end{itemize}

\section{Background and Related Work}


Increasing mobile phone usage and performance demands have led to significant research in exploring power-performance trade-offs in heterogeneous SoCs. Mobile platforms employ heterogeneous SoCs that integrate different types of cores (big/little), GPUs, and application-specific processors. They come with power management governors, 
such as on-demand~\cite{pallipadi2006ondemand}, that use simple metrics including core utilization to make power management decisions, although the increasing complexity of heterogeneous SoCs necessitates new algorithms for their dynamic resource management. 

The majority of prior power management techniques utilize control policies designed offline~\cite{vallina2012energy, pathania2014integrated}. 
For example, two recent studies propose CPU/GPU frequency selection techniques using decision tree~\cite{pathania2015power,park2017ml}. 
These techniques are designed offline and verified at runtime under gaming applications.
Purely offline methodologies are impractical since they cannot adapt to new applications with unknown characteristics.
Therefore, a practical dynamic power management technique for mobile platforms should {\em automatically} adapt to new application applications seen at runtime.


One class of techniques choose between multiple pre-designed dynamic power management policies by characterizing the incoming applications online~\cite{dhiman2009system, cochran2011pack,aalsaud2016power,singh2019collaborative}. 
For example, the technique presented in~\cite{aalsaud2016power} updates the parameters online, when a new uncharacterized application begins. 
These methods incur the cost of storing multiple control policies and switching among them online. 
The accuracy of these approaches depends critically on the availability of an appropriate policy, which is infeasible due to the rapidly growing number of new applications. 
A recent study updates the policy online with the help of a proportional-integral (PI) controller which estimates the workload behavior~\cite{won2014up}. However, this technique controls only one knob,  the uncore frequency.


\rev{Modeling power-performance with system-level measurements is usually an 
integral part of dynamic resource management techniques. Multiple studies in 
the literature discuss power modeling for heterogeneous 
platforms~\cite{sultan2014processor, brooks2007power, ye2000design}. A 
comprehensive discussion about power modeling techniques is presented 
in~\cite{sultan2014processor, brooks2007power}. One class of power models rely on the
parameters obtained through circuit simulations~\cite{ye2000design, 
bellosa2000benefits}. Since typical circuit simulations are slow, obtaining 
power models though simulations are time-consuming. Another class of techniques use empirical data to obtain the power models. Since our technique uses 
empirical data to obtain the power model, we discuss this class in more detail. 
An empirical power modeling technique is proposed 
in~\cite{bhat2017algorithmic}. In this technique, the authors collect power 
data from the platform itself by executing it at different fixed temperatures, 
voltages and frequencies. Several researchers also proposed offline and online 
performance models for heterogeneous platforms. Authors 
in~\cite{kadjo2015control, gupta2018online} proposed machine learning-based performance models 
as a function of hardware counters to estimate GPU performance. Auto-regressive 
models are used to predict the performance in heterogeneous processors both 
offline~\cite{dietrich2014lightweight} and online~\cite{dietrich2010lms}. 
Adaptive performance models for heterogeneous SoCs are proposed 
in~\cite{gupta2018staff, bhat2018online}. In these techniques, the authors 
perform offline feature selection and use Recursive Least Square (RLS) to learn 
model coefficients online. In our proposed work, we first collect power and 
execution time data for different applications. Since total power is a function 
of dynamic capacitance ($C_{dyn}$) and leakage current ($I_{leak}$), we model 
these two as a linear combination of hardware counters. However, these models 
might be inaccurate for the new applications seen at runtime. Therefore, we 
update these linear models through RLS with the help of the hardware counters 
and the power values obtained at runtime. We apply a similar technique to 
estimate execution time too.}

Recent studies proposed applying Reinforcement Learning (RL) to dynamic power management to enable online adaptation~\cite{chen2015distributed, ul2017hybrid, zhang2017energy}. 
For instance, Q-Table based approaches are used for energy-aware DVFS scheduling~\cite{chen2015distributed, ul2017hybrid}. 
These approaches are not scalable for heterogeneous SoCs since the number of state-action pairs is increasing with the number of cores and supported power states. 
To address this problem,  Zhang et al. \cite{zhang2017energy} proposed 
a deep queue learning network (DQN) technique for heterogeneous SoCs.
A common drawback of all RL methods is their inefficiency to learn the optimal policy for new applications. More specifically, RL techniques require a large number of data samples to converge, since they learn via trial-and-error using weak training signals.


\jrnAdd{In this work, we employ the principles of imitation learning (IL)~\cite{schaal1999imitation} due to significant advantages of IL methods over RL: require fewer data and time to learn near-optimal policies. IL leverages the optimal sequential-decision making behavior (referred to as the Oracle) and employs supervised learning techniques to mimic this behavior. However, exact imitation learning algorithms suffer from error propagation.
Advanced IL algorithms use additional training data collected during policy evaluation to recover from mistakes to address this problem~\cite{ross2011reduction}.  This idea has been used to control voltage-frequency islands in a multiprocessor SoC~\cite{kim2017imitation}. In their approach, the control policy is created offline using known applications and evaluation is performed for homogeneous SoCs using only simulation studies. Recently, another IL-based dynamic power management technique is proposed in~\cite{mandal2019dynamic} for heterogeneous platforms. In this approach, the authors design the Oracle policy using the applications known at design-time. Subsequently, they use the offline designed policy at runtime. Therefore, these methodologies cannot improve the control policy for new applications. 
}

In contrast to prior work, we propose a novel low-overhead \textit{Online Imitation Learning} approach that can learn to control multiple configuration knobs simultaneously as new applications arrive.
Our proposed solution utilizes power/performance models to generate strong online supervision which enables us to tailor the policy to new applications seen at runtime. Finally, we evaluate the effectiveness of our solution with experiments on a hardware platform.



\begin{table}[t]
\centering
\caption{Notations and important parameters} \label{tab:notations}
\begin{tabular}{l|l}
\hline
$\mathcal{E}$    & Sequence of epochs of an application  ($\mathcal{E} = \{E_1, E_2, \cdots, E_T \}$)                                                                                                  \\ \hline
$E_i$         & $i^{\mathrm{th}}$ epoch of an application                                                                         \\ \hline
$\mathcal{C}$ & Set of candidate configurations                                                                                             \\ \hline
$C_j$         & \begin{tabular}[c]{@{}l@{}}Configuration on which $j^{\mathrm{th}}$ epoch of an application runs\end{tabular}         \\ \hline
$n_B$         & Number of active big cores                                                                                        \\ \hline
$n_L$         & Number of active little cores                                                                                     \\ \hline
$f_B$         & Frequency of big cores                                                                                            \\ \hline
$f_L$         & Frequency of little cores                                                                                         \\ \hline
$\pi$         & \begin{tabular}[c]{@{}l@{}}A power management policy which maps $E_i$ to $C_i \in \mathcal{C}$\end{tabular} \\ \hline
\end{tabular}
\end{table}
\begin{table}[b]
	\centering
	\caption{\rev{Performance and power consumption data collected in each epoch.}}
 	\vspace{-7mm}
	\label{tab:counters}
	\begin{tabular}{@{}ll@{}}
		\\ \midrule
		Instructions Retired             & Noncache External Memory Request      \\
		CPU Cycles                       & Total Little Cluster Utilization    \\
		Branch Miss Prediction           & Per Core Big Cluster Utilization \\
		Level 2 Cache Misses             & Total Chip Power Consumption      \\
		Data Memory Access               &               \\ \bottomrule
	\end{tabular}
\end{table}
\section{Preliminaries and Problem Setup} \label{sec:prelim}

We consider a heterogeneous platform with $k$ different core types. Each type can contain 
a single core or multiple cores. For example, the platform may integrate multiple CPU cores 
while having a single GPU. Given such a platform, our goal is to design a power management 
policy to control the number of active cores and their respective frequencies. Each possible combination of the number of cores and their frequencies is a unique runtime configuration 
for the platform. Using this definition, we can represent all possible 
configurations in the system by $\mathcal{C} = \{ C_1, C_2, \cdots, C_N \}$, as 
summarized in Table~\ref{tab:notations}. For our experimental platform, 
Odroid-XU3, a configuration consists of four control knobs: the number of 
active big cores ($n_B$), the number of active little cores ($n_L$), the 
frequency of big cores ($f_B$), and the frequency of little cores ($f_L$). 
Therefore, the goal of the power management policy is to determine the tuple 
$(n_B, n_L, f_B, f_l)$ for each control interval.

The default power management governors used on mobile platforms apply decisions in periodic intervals ranging from 10--100 ms~\cite{pallipadi2006ondemand}.
These decisions typically involve determining the number of active cores and their respective frequencies. 
Suppose that we run an application repeatedly at different frequencies.
Decision intervals observed in each execution will be different from the other runs. 
This means that we cannot execute the same application at different configurations and collect the optimal configuration for each interval. Thus, periodic intervals impede construction of optimal oracles even for known applications.

\rev{
To facilitate Oracle policy construction, we segment applications into \textit{repeatable} decision epochs, which are clusters of macro-blocks. To this end, we use the methodology proposed by Gupta et al.~\cite{gupta2017dypo}. 
We start by inserting PAPI API~\cite{mucci1999papi} call within consecutive epochs using LLVM~\cite{lattner2004llvm} and clang compiler framework. Specifically, we first find all the available basic blocks in the source code. Then, we instrument at all possible basic blocks in the application. Finally, we remove PAPI calls from basic blocks at the lower hierarchy of call graph to prune the number of epochs.  As a result of the instrumentation, the number of instructions in each epoch ranges from 10 - 100 million instructions with a median of about 7.5 million instructions. This translates to about 50 - 1000 epochs per workload as a function of the total execution time of the workload.
}
This instrumentation enables us to collect the performance counters listed in Table~\ref{tab:counters} and power consumption for each \textit{repeatable} epoch.
\rev{We note that the repeatability here means that the number of instructions executed for a given epoch are always the same regardless of the configuration they are being executed on. At the same time, the other performance counters, such as LLC misses, memory accesses, can be different each time the application is executed.}
Consequently, an application with $T$ decision epochs can be expressed as a 
sequence of epochs $\mathcal{E} = \{E_1, E_2, \cdots, E_T \}$.
Each epoch has a fixed set of macro-blocks (e.g., a while loop)
ranging from 10 to 100 millions of instructions.
Using these definitions, a policy $\pi: \mathcal{E} \rightarrow \mathcal{C}^T$ is a function that maps a given epoch to one of the supported configurations. 
For instance, $\pi(E_i) = C_i$ means that epoch $E_i$ should run on configuration $C_i \in \mathcal{C}$ based on policy $\pi$.

The standard approach to obtain a policy is to first obtain data with a number of applications and then apply supervised learning methods to train a policy. However, as we show in Figure~\ref{fig:policy_pred}, policies trained offline may not perform well for unseen applications. Therefore, there is a need for methodologies that continue to learn online and adapt to unseen applications. To this end, we first present an approach using reinforcement learning to solve this problem since it is commonly used for online learning. However, RL methods require the design of a reward function and exploration of large state space. This can lead to a slow convergence for RL policies. To overcome this, we present a novel low-overhead \textit{Online Imitation Learning} approach that can learn to control multiple configuration knobs simultaneously as new applications arrive online.

\jrnAdd{\section{Canonical Baseline Approach: Reinforcement Learning (RL)}} 
%
%
%

\begin{algorithm}[b]
	\caption{Q-learning Algorithm with $\epsilon$ Greedy Exploration} \label{algo:q_learn}
\SetAlgoLined
\textbf{Input:} Number of learning epochs ($K$), Reward function ($R$), Exploration probability ($\epsilon$)\\
\For {k = $1,2, \hdots, K$}  {
 \eIf{rand() < $\epsilon$}{
    Choose $C_k$ randomly from $\mathcal{C}$ {\em // Exploration} 
 }
 { Evaluate $Q$-value for different configurations as a function of the features of the current epoch ($E_k$) \\
 Select the configuration ($C_{pol}$) with highest $Q$-value  {\em // Exploitation} \\
 Obtain reward ($R_k$) from the environment for the chosen configuration ($C_k$) \\
 Update $Q$ function following Equation~\ref{eq:dqn} \\
 $C_k = C_{pol}$}
}
\end{algorithm}

Reinforcement Learning (RL) is a commonly used framework for solving sequential decision-making problems. 
In RL setting, a controller takes an action as a function of the current system state.
In our power management problem, we define the state as the hardware counters observed in a given decision epoch. The controller takes an action as a function of the hardware counters.
After the controller takes the action, it interacts with the environment. The environment provides feedback (reward) to the controller about the quality of the action taken by the controller. The learned utility of actions at states is represented in the form of $Q$-values. Each state and action pair corresponds to a $Q$-value. Depending on the reward, the controller gets to know how good or bad the previously taken action was and the corresponding $Q$-value gets updated. Using the reward and the current state information, the controller performs an update to its parameters such that the actions in future states can be improved. After a sufficient number of observations, the controller learns to take near-optimal action for a given state. Algorithm~\ref{algo:q_learn} delineates the canonical Q-learning algorithm that is popularly employed in the design automation community.
There are different methodologies in the literature to implement this learning algorithm, especially to store $Q$-values.  One of them employs a Q-table to {\em explicitly} store the values for each state and action pair, and another one uses a function approximator to {\em implicitly} store the $Q$-values.

\vspace{1.0ex}

\noindent\textbf{Table-based approach:} A well-known methodology to implement Q-learning is the table-based approach. In this setting, the $Q$-values represent the value of taking an action in a given state. The $Q$-values of all the states and the corresponding actions are stored in a table. When the controller takes an action, the environment serves the reward using a hand-designed reward function. Subsequently, it calculates the $Q$-value for the current state-action pair and populates the table. It has been proved that table-based RL approach will asymptotically converge to take optimal actions~\cite{sutton2018reinforcement}.
Table-based RL is typically used to perform power management for SoCs in existing  literature~\cite{chen2015distributed, ge2011dynamic, shafik2015learning, ul2017hybrid}.

\vspace{1.0ex}

\noindent\textbf{Function approximation approach: } Although table-based RL 
approach is very efficient to learn an optimal controller, this approach scales 
poorly as the number of states grows beyond a small number. If there are $S$ 
possible states and $A$ candidate actions, then the number of entries in the 
Q-table is $\mathcal{O}(S \times A)$. Therefore, if $S$ is very high, then the 
number of entries in the Q-table will grow. Moreover, in the table-based 
approach, the state has to be discretized into a fixed number of bins when the 
state space is continuous, as is the case in our problem. If the discretization 
is coarse, then the learned model can suffer from low accuracy. On the other 
hand, if the discretization is very fine, then storing a large table will incur 
additional overhead. To solve these challenges with table-based policy 
representation, function approximation based Q-learning algorithm is proposed 
in the literature~\cite{mnih2015human}. In this methodology, the $Q$-values are 
learned using a function approximator, such as a neural network or regression 
trees. If a neural network is used as a function approximator, then the 
Q-learning methodology is often referred to as deep Q-Learning (DQN). Deep 
Q-Learning has been used to manage control knobs of an SoC in prior 
work~\cite{gupta2019deep, zhang2018double, tian2018multi}. We refer the reader 
to ~\cite{mnih2015human} for more details on deep Q-learning. 

The power management problem described in Section~\ref{sec:prelim} has a continuous state space and a discrete action space. Therefore, we apply Deep Q-Learning~\cite{mnih2015human} to obtain a policy $\pi$. The $Q$-values are updated following Equation~\ref{eq:dqn}:
\begin{equation} \label{eq:dqn}
    Q(E_k, C_k) = (1-\alpha)Q(E_k, C_k) + \alpha [R_k + \gamma \max_{c \in \mathcal{C}} Q(E_{k+1}, c) ]
\end{equation}
where $Q(E_k, C_k)$ is the $Q$-value for $k^{th}$ epoch executed on configuration $C_k$, $\alpha$ is the learning rate and $\gamma$ is the discount factor.

\vspace{1.0ex}

\noindent {\bf Reward function:} The performance of RL algorithms critically depends on the design of a good reward function. We define the reward function ($R_k$) at $k^{th}$  decision epoch as follows.
\begin{equation} \label{eq:reward}
    R_k = -\frac{P(E_k, c_k) \times t(E_k, C_k)^\beta}{P_{min}(E_k) \times t_{min}(E_k)^\beta}   
\end{equation}
where $P(E_k, C_k)$ and $t(E_k, C_k)$ are the power consumption and execution time respectively at decision epoch $E_k$ with configuration $C_k$. $P_{min}(E_k)$ is the minimum possible energy consumed at decision epoch $E_k$. Typically, when the powersave governor is active, energy consumption is minimum. $t_{min}(E_k)$ is the minimum possible execution time at decision epoch $E_k$. With performance governor, the platform delivers minimum execution time. Further, we can change the value of parameter $\beta$ to fine-tune the optimization objective.  For example, with $\beta = 0$, the reward function results in control behavior to minimize power, and with $\beta = 1$, the reward function results in control behavior to minimize energy. In our experiments, we employ $\beta = 1$, as this minimizes energy.
However, we argue that deep Q-Learning based RL is not suitable for controlling heterogeneous processors when compared to our proposed online imitation learning approach. This can be attributed to the following reasons: 
\begin{itemize}
    \item \textit{First,} the feedback obtained by the controller from the environment through the reward function is indirect. Therefore, the controller has to sufficiently explore the large state space to be able to learn a near-optimal policy. This can result in a long convergence time for the RL policy, as shown later in Section~\ref{sec:unseen_app}. Such a long convergence time is not desirable since the workload composition can change frequently in mobile devices. 
    \item \textit{Second,} the accuracy of policy learning in RL critically depends on the reward function. However, designing a good reward function that can achieve the desired convergence is challenging. 
\end{itemize}
In order to overcome these challenges, we propose a novel online imitation learning methodology in the next section.

\section{Online Imitation Learning Method} \label{sec:online_IL_framework}

\jrnAdd{Imitation learning (IL) is a supervised machine learning framework that aims to imitate an Oracle for sequential decision-making problems. The Oracle policy provides expert-level feedback to the policy being learned by IL. In classical IL settings, the Oracle is provided as training data from expert demonstrations (such as humans performing a task in robot learning~\cite{schaal1999imitation}). In the case of dynamic management of heterogeneous systems, the Oracle provides the best configuration for a given epoch of the application. 
The IL-based controller receives the correct control action as feedback when it makes incorrect decisions. However, RL-based controllers can only observe the reward from the environment, which is weaker supervision with respect to IL. It has been shown that IL policies converge exponentially faster than RL~\cite{sun2017deeply}. Due to this advantage, we choose IL to train our resource management policies.}

\begin{figure}[t]
	\centering
	\includegraphics[width=0.6\linewidth]{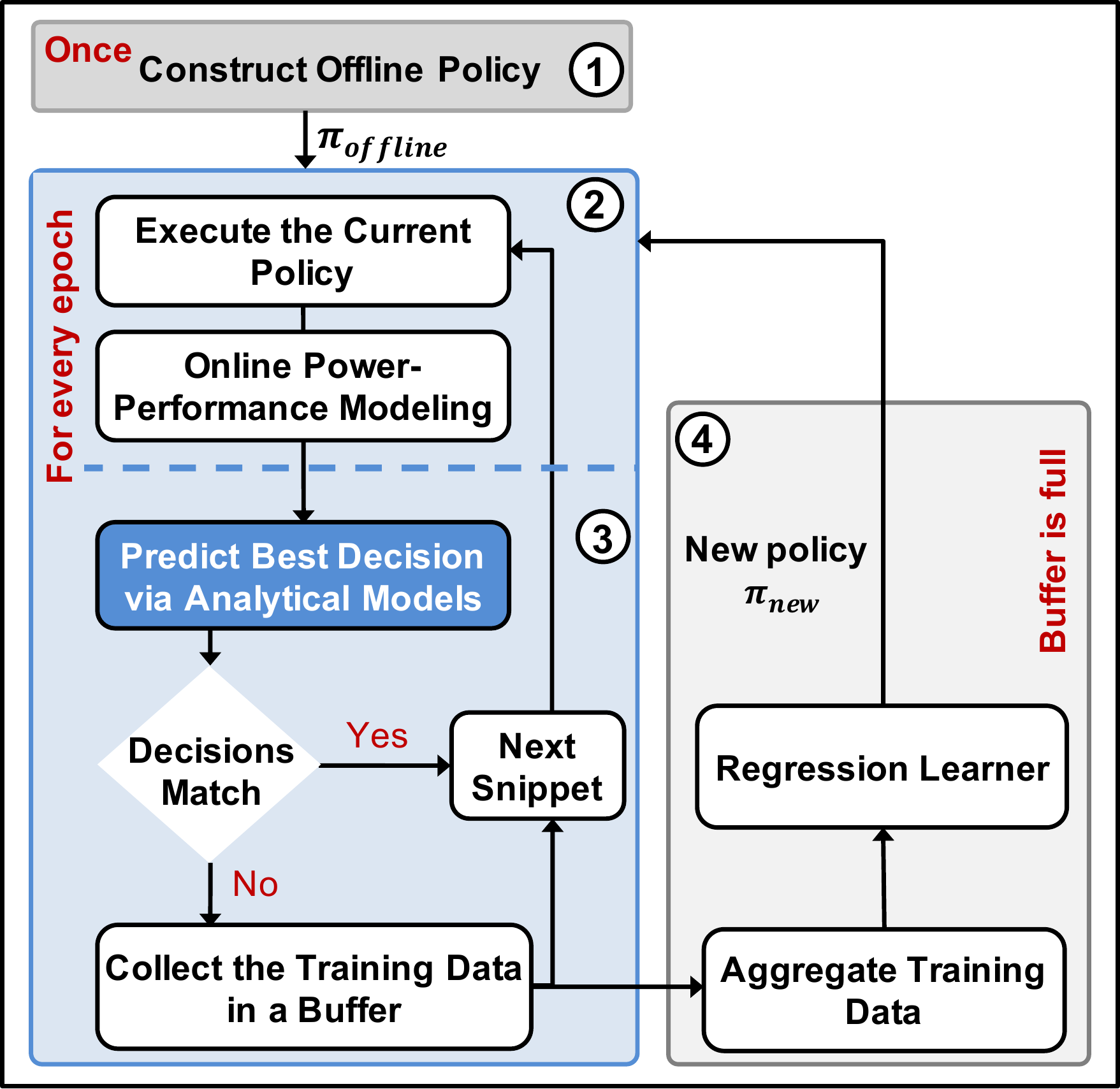}
	\vspace{-3mm}
	\caption{Overview of the proposed framework.}
	\vspace{-3mm}
	\label{fig:overview}
\end{figure}

\jrnAdd{Construction of an Oracle is one of the most important steps to train an IL policy. In an offline setting, this can be achieved by collecting characterization data and using advanced algorithms like dynamic programming to obtain the Oracle~\cite{mandal2019dynamic}.
In literature, there are techniques to construct an offline policy using IL for dynamic management of SoCs~\cite{kim2017imitation, mandal2019dynamic}.
However, constructing an Oracle policy is non-trivial for an online learning setting because of two reasons. \textit{First,} Oracle construction needs a large amount of data which is not readily available in an online setting. \textit{Second,} the methodology to construct Oracle policy may involve expensive computation which can be prohibitive at runtime. Therefore, there is a strong need for methodologies that can construct high-quality Oracle policies at runtime with minimal overhead. To address this challenge, we propose a four-step methodology for online imitation learning}
as depicted in Figure~\ref{fig:overview}. These steps are outlined below and detailed in the following sub-sections.

\vspace{1mm}
\noindent \textbf{1. Construct Control Policy Offline (Section~\ref{sec:offline_policy_construction}) --} 
The first step of the proposed methodology is to construct 
an offline control policy $\pi_{offline}$ using the set of training applications available at the design-time. Prior approaches use similar policies constructed offline to make runtime power management decisions~\cite{park2017ml, gupta2017dypo}. However, they do not learn the characteristics of new applications online and adapt the policy at runtime. In contrast, the proposed approach uses this offline policy as a starting point and employs online imitation learning to adapt it to new applications. 

\vspace{1mm}
\noindent \textbf{2. Online Execution and Power-Performance Modeling (Section~\ref{sec:online_pnp}) --}
At runtime, we use the most up to date policy trained until that point to make power management decisions. Meanwhile, we also continuously collect the workload metadata summarized in Table~\ref{tab:counters}. 
Subsequently, we employ this data for maintaining accurate runtime adaptive power-performance models with negligible overhead.

\vspace{1mm}
\noindent \textbf{3. Online Oracle Policy Construction (Section~\ref{sec:online_oracle}) --} 
The fundamental challenge in online learning is to construct a good Oracle (or reference) policy that can provide strong supervision needed to update the policy and quickly learn an optimal policy for new applications. We overcome this challenge by using our power-performance models maintained online, as shown in Figure~\ref{fig:overview}. 
More specifically, we use these models at the end of each control interval 
to determine the best action with less than 0.1\% runtime overhead.
This retrospective view enables us to compare this action against
the actual policy decision.
If they do not match, we record the metadata and optimal decision as future 
training data to update the policy. 

\vspace{1mm}
\noindent \textbf{4. Online Training (Section~\ref{sec:online_training}) --} 
After a set of metadata and corresponding optimal configuration are collected, 
we use them as supervised training examples to incrementally re-train the policy using imitation learning. Hence, the policy adapts to new applications via supervision from the oracle constructed online.

%
%

\subsection{Offline Policy Construction}\label{sec:offline_policy_construction}
We can employ any existing method to construct a control policy offline using the applications known at the design time. Without loss of generality, we use imitation learning for this purpose. 
Segmenting the applications into repeatable epochs, i.e., microbenchmarks, 
enables us to execute each epoch of an application at each supported configurations.
Hence, we collect power consumption and performance data for each epoch-configuration pair at design-time.  This data allows us to determine the configuration that optimizes a given metric (e.g., energy) for each epoch known at design-time. 
More specifically, we use this data to construct an Oracle policy $\pi^*(\cdot): \mathcal{E} \rightarrow \mathcal{C}^{T}$ that maps each epoch to the configuration that minimizes the energy consumption.

The Oracle cannot be used directly at runtime due to excessive (more than 1 MB) storage and computational requirements. Therefore, we need to construct a policy $\hat{\pi}: \mathcal{E} \rightarrow \mathcal{C}^T$ which determines the best power management decisions for each epoch $E_k$. In other words, we want the policy $\hat{\pi}$ to mimic the behavior of the Oracle policy. To this end, we apply imitation learning to first construct an offline policy using the training data for all epochs $\mathcal{E} = \{E_1, E_2, \cdots, E_T \}$ and the corresponding labels for best configuration obtained from the Oracle policy. The features for each epoch are the performance counters and power consumption values listed in Table~\ref{tab:counters}.
We use a neural network to approximate the Oracle using this training data. 
Advantages of the neural networks include the ability to learn complex decision functions and perform online updates via backpropagation. After learning a policy using exact imitation, we further improve its robustness and accuracy using state-of-the-art IL techniques~\cite{ross2011reduction} to closely approximate the behavior of Oracle policy. 
At the end of this design-time step, we obtain the offline policy $\pi_{offline}$ in the form of four functions to predict the configuration for knobs $n_B$, $n_L$, $f_B$, and $f_L$.


\subsection{Online Execution and Power-Performance Modeling}
\label{sec:online_pnp}


At runtime, we start with the offline policy ($\pi_{offline}$) created at design time. 
Subsequently, we run the incoming applications using the most up to date policy obtained so far, as shown in Figure~\ref{fig:overview}. 
Our online learning objective is to update the parameters of policy to minimize the overall energy consumption. Meeting this goal requires a reliable supervision, but constructing an Oracle similar to one presented in Section~\ref{sec:offline_policy_construction} is not practical at runtime. 
Therefore, we also collect power consumption and performance data listed in Table~\ref{tab:counters} and maintain accurate analytical power-performance models during the regular execution at runtime.
Then, we use these models to achieve the online learning objective.

\begin{figure}[t]
	\centering
	\includegraphics[width=0.6\linewidth]{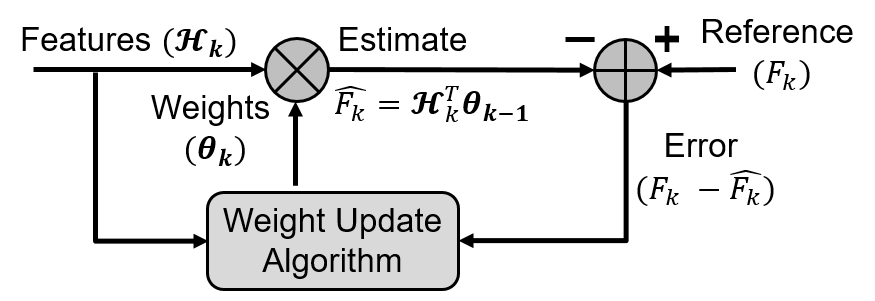}
	\vspace{-4mm}
	\caption{Illustration of the online recursive least square algorithm to update the analytical models of power consumption and execution time.}
	\label{fig:RLS}
	\vspace{-1mm}
\end{figure}
\begin{algorithm}[b]
	\caption{Online power/performance modeling} \label{alg:pnp_model}
\SetAlgoLined
\textbf{Input:} Features listed in Table~\ref{tab:counters} ($\mathbf{{h_k}}$),  Measured power/performance ($F_k$)\\
\For {each epoch $E_k$}  {
 Obtain weights at the end of the previous epoch ($\mathbf{\theta_{k-1}}$) \\
 Estimate power/performance through a linear model: $\widehat{F_k} = \mathbf{h_k^T} \mathbf{\theta_{k-1}}$ \\
 Calculate the error between the measured value and the estimated value: $F_k - \widehat{F_k}$ \\
 Apply weight update algorithm to obtain weights ($\mathbf{\theta_{k}}$) for the next epoch.
}
\end{algorithm}

We begin with the power consumption and execution time models built offline with the help of the same applications used for the design-time Oracle construction.
For example, the total power in $k^\mathrm{th}$ epoch is expressed as:
\begin{align} \label{eq:cpu_power} \nonumber
P_{CPU}  & = P_{dyn} + P_{leak} \\
& = \bigg( C_{dyn}(\mathbf{h_k}, \mathbf{\theta_k})V^2f \bigg)  + VI_{leak}(\mathbf{h_k}, \mathbf{\theta_k})
\end{align}
%
%
where the dynamic switching capacitance $C_{dyn}$ and leakage current $I_{leak}$ are modeled as a function of the performance counters ($\mathbf{h_k}$) shown in Table~\ref{tab:counters}. 
We employ a linear model using these counters and weight vector $\mathbf{\theta_k}$, 
as depicted in Figure~\ref{fig:RLS}. 
Note that our data also includes measured power consumption enabled by current 
sensors~\cite{ODROID_Platforms}. 
At the end of each epoch, we employ our model to estimate the power consumption in the previous interval. 
Subsequently, we compare it to the actual measurement, as illustrated in Figure~\ref{fig:RLS} and Algorithm~\ref{alg:pnp_model}.  We feed this error to the Recursive Least Square (RLS)~\cite{mendel1995lessons} algorithm 
to update the parameters of the power consumption model. Similarly, we maintain an execution time model as a function of the performance counters. In this case, the reference is simply the length of the epoch, which can be tracked easily at runtime.

In summary, we maintain accurate power-performance models by using the data available at the end of each epoch. These models are used to generate strong supervision required by our online learning algorithm as described in the following section. We also note that updating the power and performance models have less than 0.1\% overhead with respect to the execution time of an epoch.

\subsection{Online Oracle Construction} \label{sec:online_oracle}

In general, providing online supervision in terms of the best configuration to update the policy is a challenging problem. We overcome this challenge by leveraging the power and performance models, as described in the previous section. 
At the end of each epoch ($E_k$), we have the metadata listed in Table~\ref{tab:counters} and up to date power-performance models. These models and the inputs enable us to evaluate the energy at candidate configurations in $\mathcal{C}$ other than the one chosen by the current policy. Specifically, we use the hardware counters measured on the current configuration to estimate the power and performance when using other configurations.
\rev{In general, the hardware counters do not remain the same when we estimate the power consumption of other configurations. Indeed, predicting the change in the counters is a complex problem which involves modeling of system dynamics and the state transition probability. This can potentially add additional overhead to the Oracle estimation step of our approach. Therefore, to avoid additional overhead, we reuse the hardware counter values that we observe at runtime (of the current configuration) to estimate the power consumption of other configurations.}
After obtaining the energy consumption values at candidate configurations in $\mathcal{C}$, we mark the configuration consuming the least energy as the optimal configuration ($C_k^*$).

We note that exhaustive search in the whole configuration space is resource expensive. 
For example, Samsung Exynos 5422 processor supports up to 4940 different configurations. 
Our implementation shows that evaluating each configuration can take up to 5\%  of the execution time of an epoch, which is prohibitive at runtime. 
To overcome this difficulty, we exploit the fact that the current policy ($\hat{\pi}$) is reasonably accurate and makes decisions that are not too far from those made by the Oracle policy.  Therefore, instead of searching in the whole configuration space, 
we search within a local neighborhood of the configuration predicted by the current policy as described in Algorithm~\ref{alg:oracle_algorithm}. 

\vspace{1.0mm}

\noindent\textbf{Resource-bounded Online Oracle Construction:} We start the local search in the neighborhood of $\hat{C_k}$ by setting a maximum allowable time limit for the search, such as 5 $\mu s$. 
We initialize the optimal configuration $C_k^*$ to the current policy decision ($\hat{C_k}$) and optimal cost ($J_k^*$) to the cost of $\hat{C_k}$ (line 3 in Algorithm~\ref{alg:oracle_algorithm}). After this step, we start performing a search around the configuration chosen by the current policy. We first evaluate the metrics obtained by configurations that are one step away from the policy decision. 
Specifically, we get configurations to be evaluated by changing the frequency of the cores by 200 MHz (i.e., $\pm$ 200 MHz) or changing the number of active cores by 1 (i.e., $\pm$ 1). After evaluating all the configurations that are one step away, we find the configuration with the minimum cost from $\mathcal{N}(C)$ as shown in Algorithm~\ref{alg:oracle_algorithm}.
We update the optimal configuration ($C_k^*$) whenever a configuration in the neighborhood search outperforms the best configuration found so far~(line 10 in Algorithm~\ref{alg:oracle_algorithm}). If the configuration chosen by the current policy is different than $C_k^*$, then we store $C_k^*$ along with the features of the system state in a buffer, as illustrated in Figure~\ref{fig:overview}. We continue the search in this way until the time limit is exceeded. The data in this buffer serves as the supervised training data from Oracle which is used for online learning, as detailed in the next section.
This flexible methodology can be easily extended to different objectives. 
For example, we can perform a gradient-based search in the direction of increasing value if the objective function is to maximize performance.

\begin{algorithm}[t]
	\caption{\jrnAdd{Online Estimation of Oracle Configuration}} \label{alg:oracle_algorithm}
\SetAlgoLined
\textbf{Input:} Power and performance models; $\hat{C_k}$, predicted configuration; \textit{Timeout period} \\
\textbf{Output:} Oracle configuration ($C_k^*$) at decision epoch $E_k$\\
Initialize: $C_k^* \gets \hat{C_k}$, $C_{min} \gets \hat{C_k}$, $J_k^* \gets J(\hat{C_k})$\\
\While{!Timeout}{
$\mathcal{N}(C)$ $\gets$ Configurations with control knob values one step away from $C_{min}$ (e.g. ${n_B} \pm 1$)\\
Evaluate cost of configurations in $\mathcal{N}(C)$  \\
$C_{min} \gets $ Configuration with minimum cost in  $\mathcal{N}(C)$  // Greedy search step \\
$J_{min} \gets J(C_{min})$ \\
\If{($J_{min} < J_k^*$)}{
$J_k^* \gets J_{min}$ \\
$C_k^* \gets C_{min}$
} 
}
\end{algorithm}

\begin{algorithm}[b]
	\caption{Policy Update via Online Training} \label{alg:online_training}
\SetAlgoLined
\textbf{Input:} Current Policy $\pi$, Online Oracle $\pi^*$ via power-performance models\\
\For {each epoch $E_k$}  {
 \If {$\pi(E_k)$ != $\pi^*(E_k)$} {
  Record hardware counters and the Oracle configuration $\pi^*(E_k)$ as a new training example in the buffer
 }
 \If {Buffer is full} {
  Perform online training to update the parameters of policy (i.e., neural network) via backpropagation algorithm using the new training examples \\
  {\bf return} the new policy $\pi_{new}$
 }
}
\end{algorithm}
\subsection{Policy Update via Online Training} \label{sec:online_training}
The final step in the proposed online-IL methodology is to incrementally update the parameters of the policy as a function of the new unseen applications and new training examples. 
The online training method to update the policy is shown in Algorithm~\ref{alg:online_training}. 
As described in the previous section, 
the Oracle policy (i.e., $\pi^*(E_k) = C_k^*$) is stored in a buffer. 
The size of this buffer is important since it determines the training accuracy and implementation overhead. 
Our experimental evaluations show that metadata and label for 100 epochs easily provide close to 100\% accuracy in adapting to new applications. 
Furthermore, the corresponding storage overhead is less than 20KB. 
Therefore, we use a buffer to store 100 Oracle entries in our experiments. 
We aggregate the training examples until the buffer is full. 
\rev{Subsequently, we update the parameters of the policy, which is represented as a neural network using this training data and backpropagation algorithm~\cite{hecht1992theory}.  
%
Once the online training procedure to update the policy parameters is completed, we replace the current policy with the new policy $\pi_{new}$ and continue the next iteration of online-IL approach.}


\section{Experimental Evaluation} \label{sec:experimental_eval}

\subsection{\jrnAdd{Experimental Setup}}
\noindent\textbf{Platform:} We implement the proposed online IL algorithm on Odroid-XU3~\cite{ODROID_Platforms} board running Ubuntu 15.04 OS. The board integrates a Samsung Exynos 5422 SoC that consists of a A15~(big) core cluster with 4 cores, a A7~(little) cluster with 4 cores, ARM Mali GPU and other components. In order to facilitate power measurements, the board also provides sensors to measure the power consumption of the big cores, little cores, main memory, and the GPU.
We sample these sensors every 50~ms to record the power consumption of each component. These measurements are used in the online learning algorithm to update the power/performance models and to evaluate the performance of the proposed online learning algorithm. 
The constructed policies are implemented as user-space governors on the Odroid-XU3 board. Specifically, the frequency of operation is specified in a \textit{sysfs} entry that is read by the kernel to set the frequency. The number of cores for the little and big cluster are set using the dynamic hotplugging feature of the Linux kernel.

\noindent\textbf{Benchmarks}: We evaluate the proposed methodology on a total of 16 applications from Mibench~\cite{guthaus2001mibench}, 
Cortex~\cite{thomas2014cortexsuite}, and PARSEC~\cite{bienia2008parsec} benchmark suites. 
\minrev{Applications from these benchmark suites exhibit a wide range of characteristics, such as compute intensiveness, memory intensiveness, and parallelism. Therefore, these applications represent a broad group of applications commonly executed in mobile platforms.}
We divide the benchmarks into two sets, as shown in Table~\ref{tab:benchmark-list}. At any time, one of the sets is used for the offline learning phase while the other set is reserved for online learning. 

\begin{table}[b]
\vspace{-3mm}
	\centering
	\caption{\jrnAdd{List of applications instrumented from different benchmark suites.}}
	\label{tab:benchmark-list}
	\begin{tabular}{@{}ll@{}}
		\toprule
		\textbf{Set 1}    & \textbf{Set 2} \\
		\midrule 
BML  & AES \\
Dijkstra         & Kmeans \\
FFT              & Spectral                  \\
Patricia         & Motion Estimation                  \\
Qsort            & PCA                                              \\
SHA              & Blackscholes-2T                                  \\
Blowfish         & Blackscholes-4T                                  \\
String Search    &                                                  \\
ADPCM            &                                                  \\
              \bottomrule
	\end{tabular}
\end{table}

\noindent\textbf{Data Collection:} Construction of an Oracle for offline policy 
requires characterization of the applications while running at different 
configurations. Therefore, we perform extensive data collection on the 
Odroid-XU3 platform while running the benchmarks. Specifically, we sweep the 
core configuration from \textit{1 big--1 little} to \textit{4 big--4 little}. 
Within each configuration, we change the frequency of big cores from 
600~MHz--2.0~GHz and little cores from 600~MHz--1.4~GHz in steps of 200~MHz. We 
do not go below 600~MHz as lower frequencies do not provide better 
energy-efficiency~\cite{aalsaud2016power}. This characterization data is used 
to construct the Oracle policies in the offline training phase.

\begin{table}[t]
\centering
\caption{Different Parameters for Neural Network } \label{tab:hyp_train_nn}
\vspace{-3mm}
\begin{tabular}{|c|l|l|}
\hline
\multirow{7}{*}{\begin{tabular}[c]{@{}c@{}}Model Hyper-\\ Parameters\end{tabular}}                                        & No. of Hidden Layers & 2                                                                   \\ \cline{2-3} 
                                                                                                                          & No. of Neurons       & 20 in each layer                                                    \\ \cline{2-3} 
                                                                                                                          & Activation           & ReLu                                                                \\ \cline{2-3} 
                                                                                                                           
                                                                                                                          & Optimizer            & Adam                                                                \\ \cline{2-3} 
                                                                                                                          & Learning Rate        & 0.001                                                               \\ \cline{2-3} 
                                                                                                                          & Loss Function        & \begin{tabular}[c]{@{}l@{}}Categorical\\ Cross-entropy\end{tabular} \\ \hline
\multicolumn{1}{|l|}{\multirow{2}{*}{\begin{tabular}[c]{@{}l@{}}Training Parameters\\ for Offline Learning\end{tabular}}} & Batch Size           & 150                                                                 \\ \cline{2-3} 
\multicolumn{1}{|l|}{}                                                                                                    & Epochs               & 500                                                                 \\ \hline
\multicolumn{1}{|l|}{\multirow{2}{*}{\begin{tabular}[c]{@{}l@{}}Training Parameters\\ for Online Learning\end{tabular}}}  & Buffer Size           & 100                                                                  \\ \cline{2-3} 
\multicolumn{1}{|l|}{}                                                                                                    & Batch Size               & 20                                                                  \\ \cline{2-3}     & Epochs        & 20                \\ \hline
\end{tabular}
\vspace{-3mm}
\end{table}
\subsection{Evaluation of the Policy Trained Offline}\label{sec:offline_policy}
To obtain a baseline, we first design the initial offline policy using only the applications known at design-time. 
We employ IL to construct the policy using a neural network with two hidden layers. 
The parameters of the policy and the hyperparameters used in training are listed in Table~\ref{tab:hyp_train_nn}.
We adopt a neural network implementation over other regression techniques, 
such as linear regression and regression trees, since neural networks allow us to learn non-linear decision functions and facilitate online updates through the backpropagation algorithm~\cite{hecht1992theory}. 
We use a smaller number of training epochs and smaller batch size for online learning because the policy updates at runtime are incremental with a relatively small number of training examples.
We observe that a training buffer size of 100, and 20 epochs with a batch size of 20 are sufficient to update the policy.

We evaluate the accuracy of the policy for each configuration knob by measuring the distance between the policy decision and a golden reference, 
which is constructed to evaluate the proposed policies.
If the total number of available levels for a configuration knob is $L$, level chosen by the policy is $L_{\pi}$ and the reference level is $L_{ref}$, then accuracy can be expressed as:
\begin{equation}\label{eq:config_accuracy}
\mathrm{Accuracy}(\%) = 100 \times \bigg(1 - \frac{|L_{\pi}-L_{ref}|}{L-1}\bigg)
\end{equation}
\begin{figure}[t]
	\centering
	\includegraphics[width=0.7\linewidth]{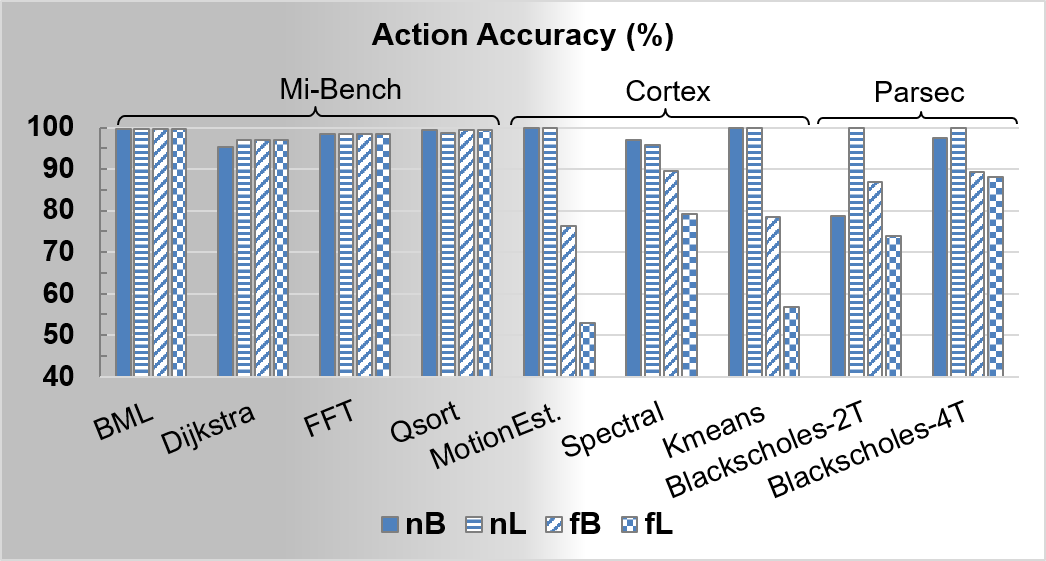}
	\caption{Accuracy comparison for all knobs for applications from Mibench, Cortex and PARSEC benchmark suites \textit{when the policy is trained with only Mibench applications.}}
	\label{fig:config_accur_offline}
\end{figure}
\begin{figure}[t]
	\centering
	\includegraphics[width=0.7\linewidth]{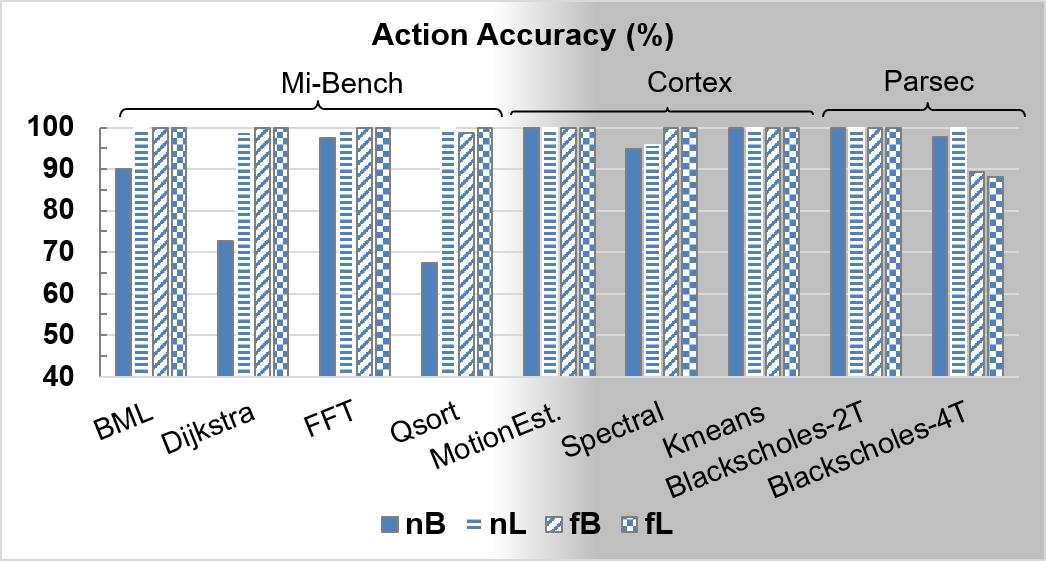}
	\caption{Accuracy comparison for all knobs for applications from Mibench, Cortex and PARSEC benchmark suites \textit{when the policy is trained with Cortex and PARSEC applications.}}
	\label{fig:config_accur_offline_cortex}
\end{figure}
After finding the accuracy of each epoch, we take the average across all epochs of an application to compute the overall accuracy for each control knob.
We note that 100\% accuracy denotes that the decisions of the underlying policy 
match with the golden reference, and no policy can perform better than that.
We first train the policy offline using applications from the Mibench suite. 
Then, we evaluate the trained policy with applications from all three benchmark suites, as shown in Figure~\ref{fig:config_accur_offline}.
Since the Mibench applications are included in the training set for this experiment, 
the offline policy always reaches more than 95\% accuracy compared to the Oracle for all four configurations knobs.
However, the accuracy is significantly lower for the applications from the Cortex suite which are not included in the training set.
For example, MotionEstimation shows only 77\% and 53\% accuracy for big and little core frequencies, respectively. 
Similarly, the accuracies for the Blackscholes application from the PARSEC suite are lower than 80\% for the number of big cores and frequency of little cores. 
\textit{This shows that the offline policy is not good enough to capture the characteristics of applications not seen at design-time.}

We also construct an offline policy by using only the  applications from Cortex and PARSEC suites.
This time the offline policy performs well when running applications from these two benchmark suites, 
but it shows poor performance for Mibench suite, as shown in 
Figure~\ref{fig:config_accur_offline_cortex}.
More precisely, we observe above 95\% accuracy for all knobs when running applications from Cortex and PARSEC suites. 
In contrast, the accuracy for applications from the Mibench suite are significantly lower. 
For example, Qsort has only 67.5\% accuracy for number of big cores.
\textit{The degradation of accuracy with unseen applications demonstrates the need for online learning.}

\begin{figure}[t]
	\centering
	\includegraphics[width=1\linewidth]{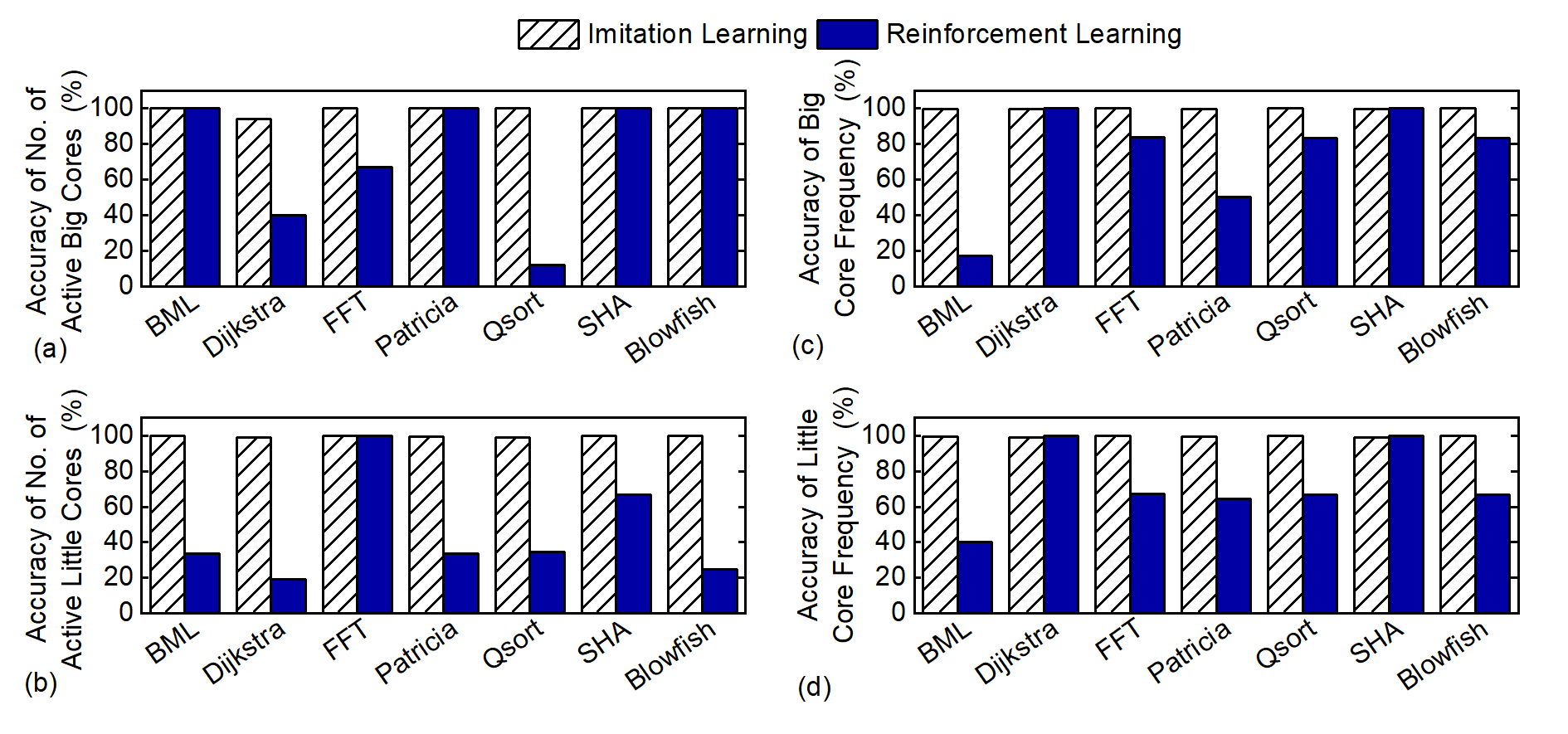}
	\vspace{-7mm}
	\caption{\jrnAdd{Accuracy comparison \textit{when the policy is trained with only Mibench applications (Set 1 in Table~\ref{tab:benchmark-list}).}}}
	\label{fig:offline_accuracy_comp}
\end{figure}

\jrnAdd{Next, we compare the performance of IL and RL policies that are trained offline. Figure~\ref{fig:offline_accuracy_comp} shows the accuracy of the offline policies trained with IL and RL, respectively.
In this case, we construct two offline policies through IL and RL by considering all 9 applications from Mibench suite as training set.
Then, we test these policies individually on all applications in Mibench suite. 
We observe that the IL policy achieves close to 100\% accuracy for all the control knobs for all applications.
That means IL policy can achieve performance similar to that of the Oracle policy.
However, the offline RL policy has much lower accuracy when compared to the IL policy.
For example, number of big cores has 17\% accuracy and number of little cores has 33\% accuracy for Qsort application.
This shows that even in the offline setting, the IL policy achieves a superior performance when compared to the RL policy.
Moreover, the offline RL policy requires more than 10$\times$ time to converge than the offline IL policy.}

%
%
\begin{figure}[b]
\vspace{-3mm}
	\includegraphics[width=0.6\linewidth]{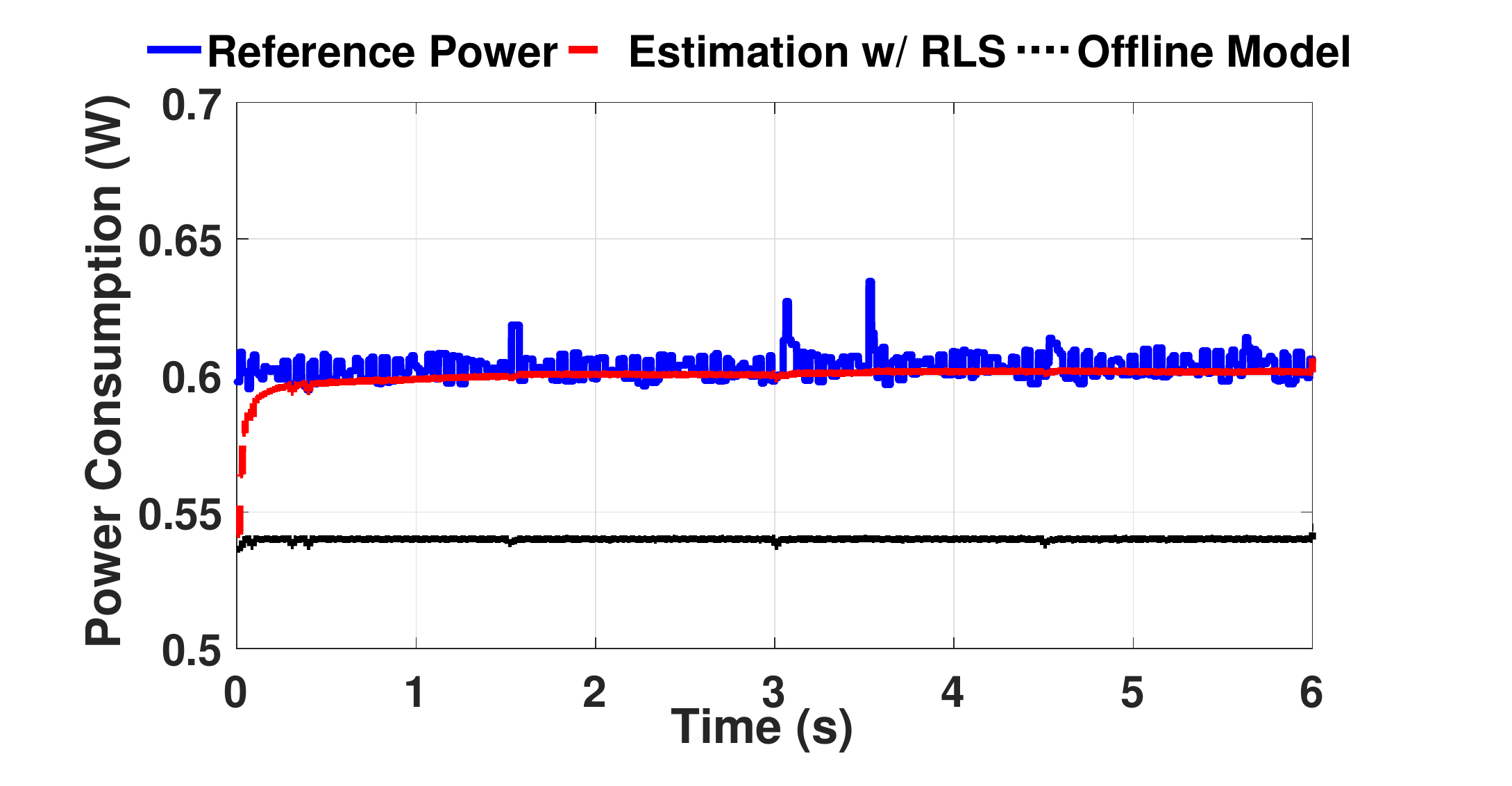}
	\vspace{-5mm}
	\caption{Comparison of estimated power for Kmeans application. While estimating with RLS, the power estimation accuracy increases from 90\% to 99\% within 0.6 second. }
	\label{fig:power_model_comp}
	\vspace{-3mm}
\end{figure}

\begin{figure*}[t]
    \begin{subfigure}{0.48\textwidth}
	\centering
	\resizebox{1.0\columnwidth}{!}{\includegraphics[width=0.8\textwidth]{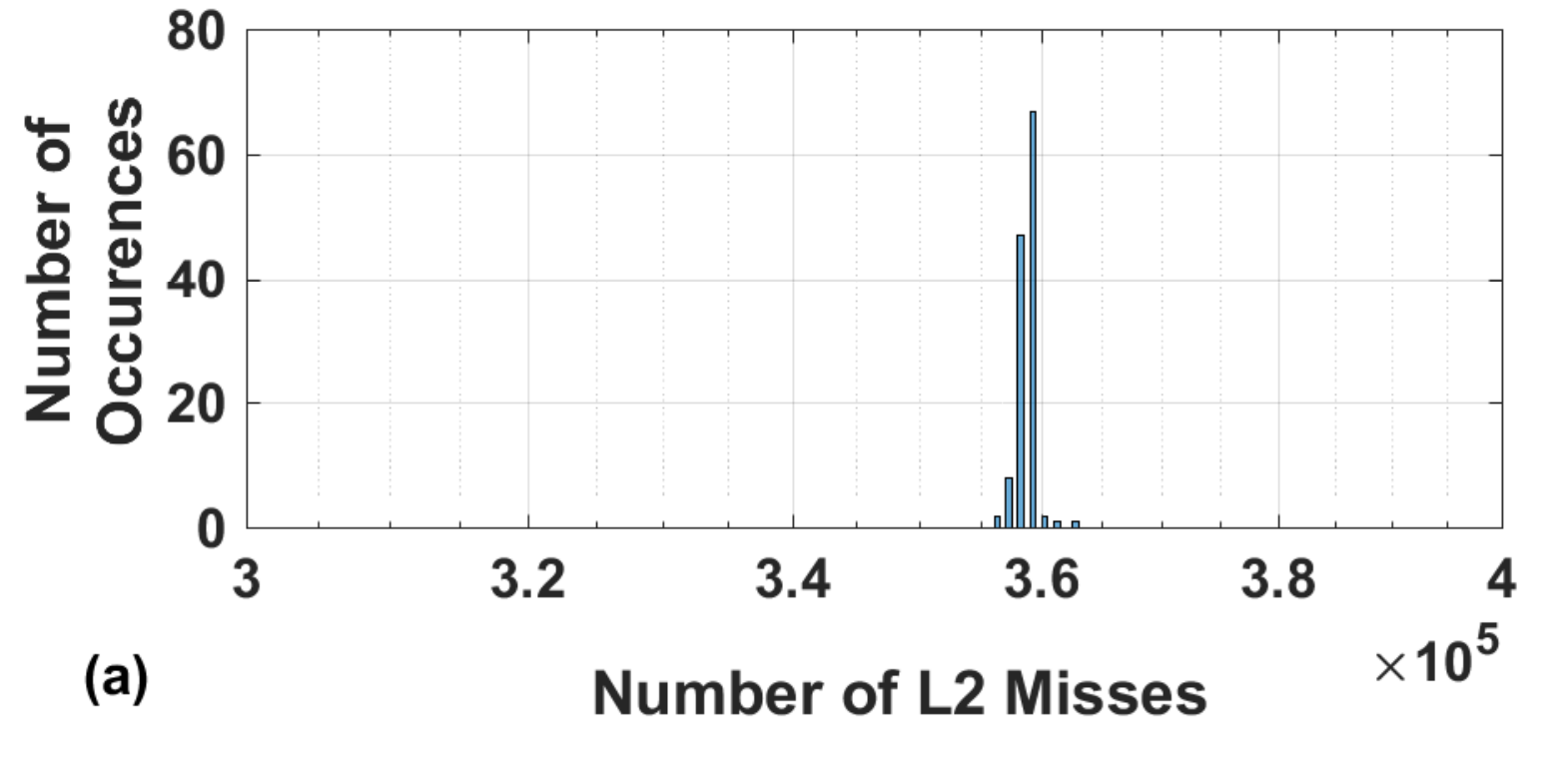}}
	\vspace{-3mm}
	\end{subfigure} \hfill
    \begin{subfigure}{0.48\textwidth}
	\centering
	\resizebox{1.0\columnwidth}{!}{\includegraphics[width=0.8\textwidth]{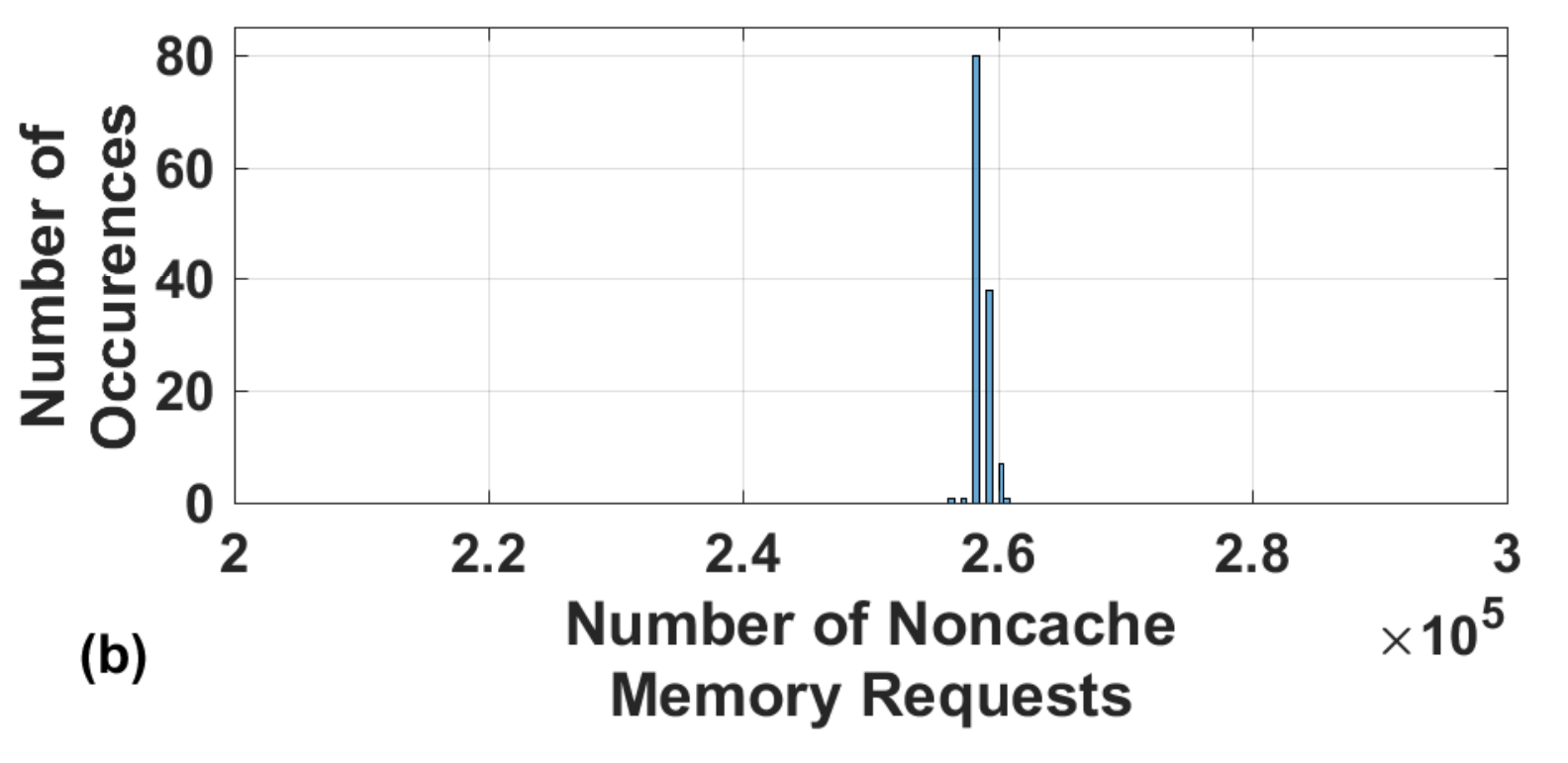}}
	\vspace{-3mm}
	\end{subfigure}
	\caption{Histogram showing (a) number of L2 misses and (b) number of noncache memory request for a particular epoch of Spectral application.}
	\label{fig:hist_counters}
\end{figure*}

\subsection{Evaluation of Power-Performance Models} \label{sec:power_exp}

\rev{Similar to the offline policy, 
we construct offline power and performance models 
using the applications available at design-time.}
These power and performance models may not be valid for 
new applications encountered only at runtime
if they are not represented by the training data. 
For example, Figure~\ref{fig:power_model_comp} 
shows that power models constructed using Mibench applications cannot track the power consumption 
while running Kmeans application from the Cortex suite. 
In contrast, our online power models based on RLS 
converge to the measured power in 
about 200~ms, as depicted by the dashed line in Figure~\ref{fig:power_model_comp}. 
Although the Kmeans application is not seen during training, 
we achieve less than 1\% error compared to the reference value. 
Similarly, the performance models converge to reference
values in 100-300~ms, which amounts to less than 10 epochs.
This shows that the proposed technique adapts the offline model to unseen applications at runtime with high efficiency.
\jrnAdd{Overall, the power models have an average error of 2.6\%.}
These models enable us to construct the accurate runtime supervision required for adapting the offline policy.
From our experiments, we observe that generating runtime supervision and updating the power and performance models at each epoch take approximately 8.5~$\mu$s.
This is negligible compared to the execution time of an epoch which is in the order of 10 ms.

\noindent\textbf{Accuracy of Power Estimation Across Configurations:} We use the performance counters measured at a given configuration to estimate the power consumption at other configurations.
\minrev{ Figure~\ref{fig:hist_counters}(a) shows the histogram of L2 misses for different configurations for a particular epoch of Spectral application. The number of L2 misses for the applications we consider is in the range of 500-2.5$\times$10$^8$. However, the number of L2 misses (for the particular epoch of Spectral application) of 122 configurations out of total 128 configurations (96\%) is between 3.57$\times$10$^5$-3.59$\times$10$^5$, i.e. within 0.6\% of each other. Therefore, the number of L2 misses is nearly the same for all configurations for that particular epoch. The histogram for the number of non-cache memory requests is shown in Figure~\ref{fig:hist_counters}(b). The number of non-cache memory requests for the applications we consider is in the range of 20-7.6$\times$10$^7$. However, The number of non-cache memory requests (for the particular epoch of Spectral application) of 98\% of all configurations is between 2.58$\times$10$^5$-2.6$\times$10$^5$, i.e. within 0.8\% of each other. We observe similar behavior for other hardware counters and for other applications too. For this reason, we can safely take the hardware counters of the current configuration as the features to estimate the power of other configurations. Since we also consider the configuration itself as a feature to estimate the power/performance of the configuration, we obtain different power/performance estimation for different configurations.}
To validate the accuracy of this power estimation, we compared the estimated power of all the configurations with the actual power consumption of the configuration. That is, we use the hardware counters observed for a single configuration to estimate the power consumption of all the configurations and compared it with the measured value. For the applications used in online training, we observed that the estimation error is around 10\%. This error is acceptable since the online policy converges to the optimal for unseen applications.

\subsection{Evaluation on Stream of Unseen Applications} \label{sec:unseen_app}

In this section, we evaluate the effectiveness of the 
proposed online-IL methodology to new applications 
observed only at runtime. 
We start with a policy which is trained offline with applications from Mibench suite.
Then, we run a random sequence of seven applications from the Cortex and PARSEC suites (AES, Kmeans, Spectral, Motion Estimation, PCA, 2-threaded Blackscholes, and 4-threaded Blackscholes) back to back.
Each complete run of this set of applications takes around 33 seconds.
We repeat this experiment for 5 different random sequences.
\begin{figure*}[b]
	\centering
	\includegraphics[width=1\linewidth]{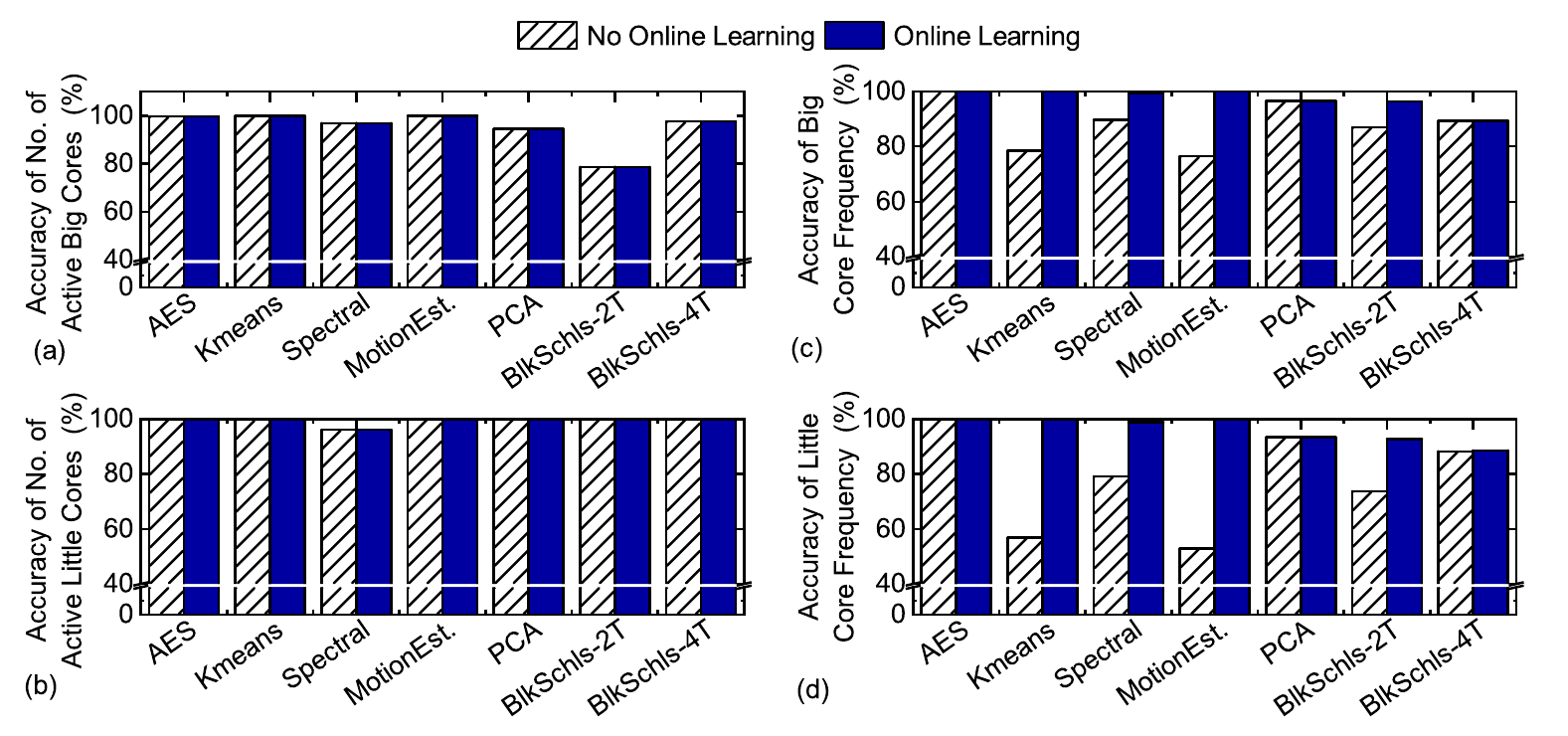}
	\caption{Comparison of accuracy of (a) number of big cores, (b) number of little cores, (c) big core frequency, and (d) little core frequency for applications in Cortex and PARSEC benchmark suites.}
	\label{fig:accur_all}
\end{figure*}
After running the application sequences, we evaluate the effectiveness of the proposed approach on each application individually.
To achieve this, we average the accuracy of each control knob over all random sequences.
We note that here accuracy measures the distance of the policy decision to the golden reference (Equation~\ref{eq:config_accuracy}).
Figure~\ref{fig:accur_all} compares the accuracy between the offline policy and the online policy after it converges for all four configurations for the random sequences.
Figure~\ref{fig:accur_all}(a) and Figure~\ref{fig:accur_all}(b) show that 
our offline policy provides good accuracy for the number of cores. 
The online policy maintains this accuracy without providing significant improvement. 
In contrast, Figure~\ref{fig:accur_all}(c) and Figure~\ref{fig:accur_all}(d) show that the
proposed online learning technique achieves a significant increase in the accuracy of little and big core frequencies. 
The biggest improvements (20\%--47\%) are observed for Kmeans, Spectral, and MotionEstimation applications.
In particular, the action accuracy of the little core frequency for MotionEstimation application increases from 53\% to 99\%. 


\begin{figure}[t]
	\centering
	\includegraphics[width=1\linewidth]{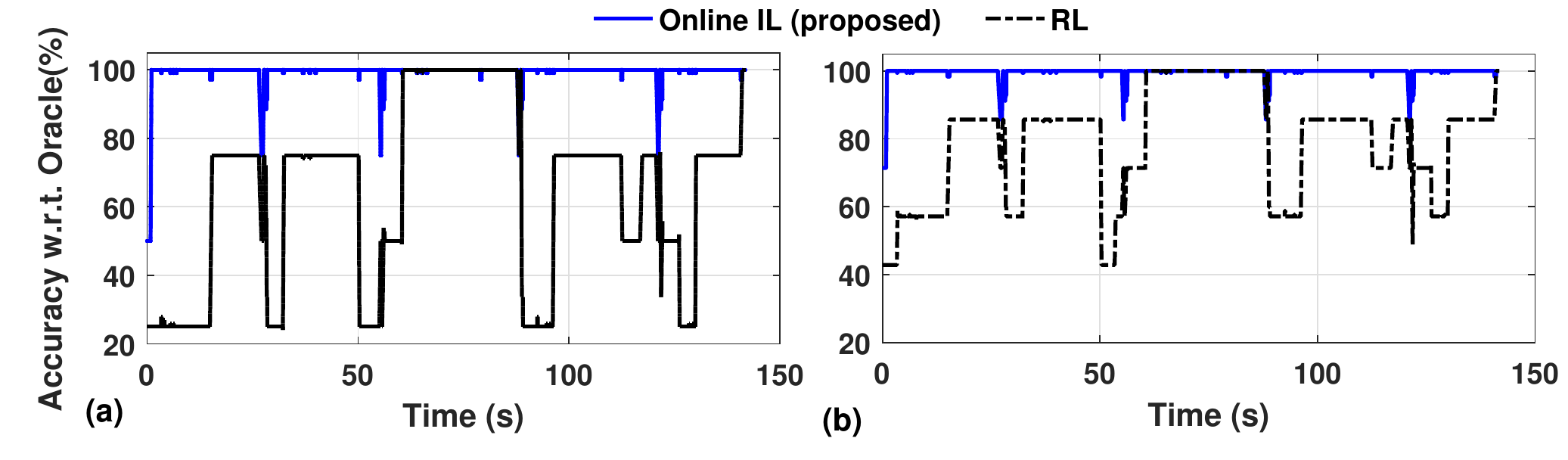}
	\caption{Comparison between proposed online IL and reinforcement learning (RL) with application sequence 1 for (a) frequency of big cores ($f_B$), (b) frequency of little cores ($f_L$)}
	\label{fig:il_rl_w_off_seq1}
	\centering
	\includegraphics[width=1\linewidth]{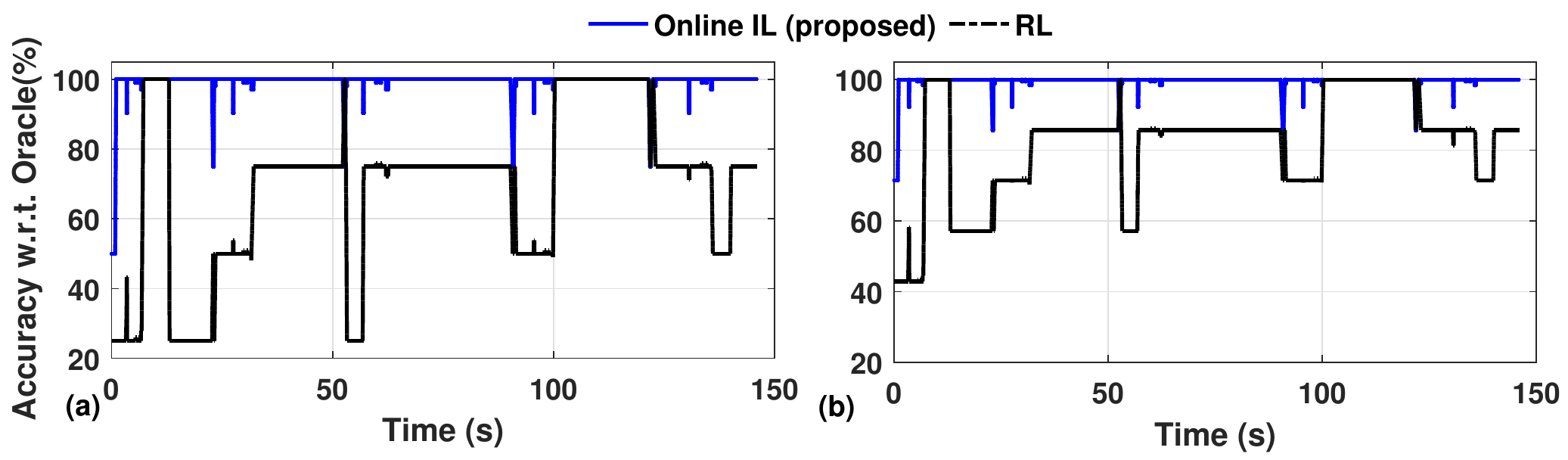}
	\caption{Comparison between proposed online IL and reinforcement learning (RL) with application sequence 2 for (a) frequency of big cores ($f_B$), (b) frequency of little cores ($f_L$)}
	\label{fig:il_rl_w_off_seq2}
\end{figure}

\begin{figure*}[t]
    \begin{minipage}{0.48\textwidth}
	\centering
	\resizebox{1.0\columnwidth}{!}{\includegraphics[width=0.8\textwidth]{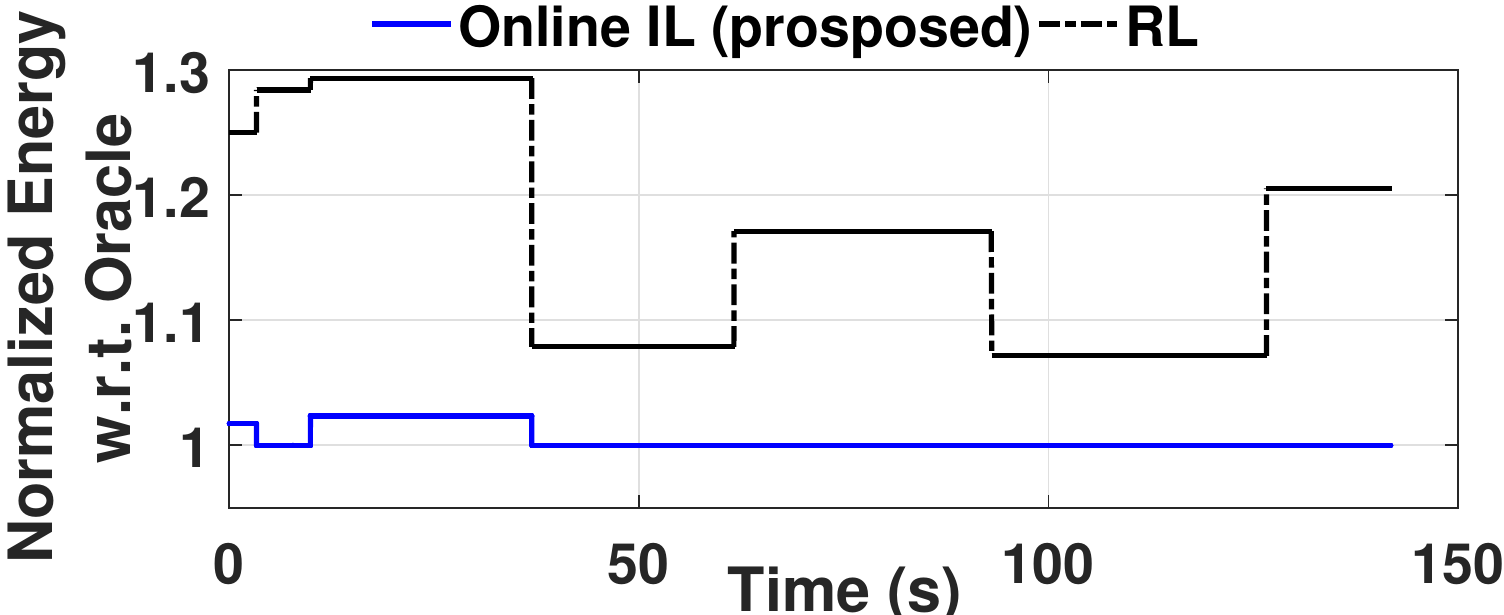}}
	\vspace{-3mm}
	\caption{Energy comparison with application sequence 1.}
	\label{fig:energy_set_1_w_off}
	\end{minipage} \hfill
    \begin{minipage}{0.48\textwidth}
	\centering
	\resizebox{1.0\columnwidth}{!}{\includegraphics[width=0.8\textwidth]{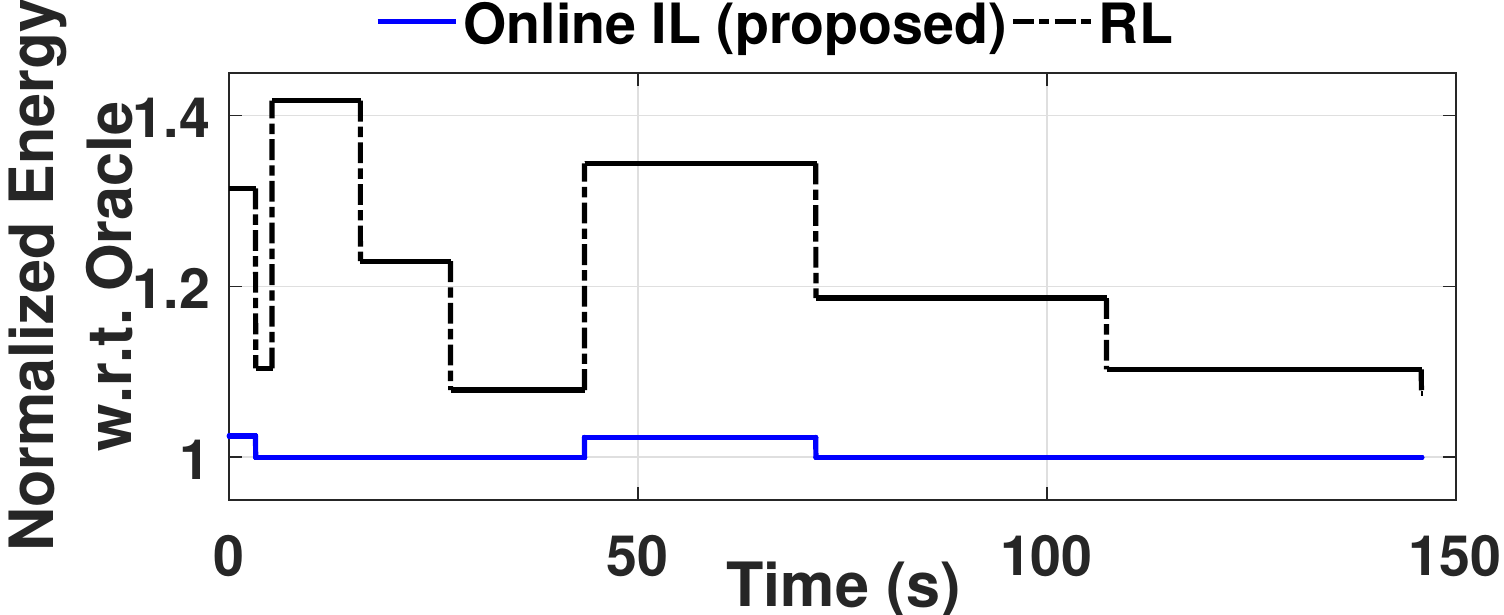}}
	\vspace{-3mm}
	\caption{Energy comparison with application sequence 2.}
	\label{fig:energy_set_2_w_off}
	\end{minipage}
\end{figure*}

We also analyze how quickly the online policy converges to the highest accuracy for different application sequences.
As stated before, we randomly construct five different application sequences.
To illustrate the convergence properties, we pick two representative application sequences.
\textit{Sequence 1} consists of Kmeans, AES, Spectral, MotionEstimation, PCA, BlackScholes-2T, and BlackScholes-4T applications one after another.
Figure~\ref{fig:il_rl_w_off_seq1}(a) and Figure~\ref{fig:il_rl_w_off_seq1}(b) shows that the accuracy of big ($f_B$) and little ($f_L$) core frequencies start at 52\% and 71\%, respectively. 
Our online-IL technique increases both of these accuracies to 100\% within 0.9 seconds,
i.e., after running less than 3\% of the sequence.
\jrnAdd{Moreover, we also use the offline RL policy to perform online learning for this application sequence.
To perform experiment with RL, we take the offline RL model trained with applications from Mibench suite as described in Section~\ref{sec:offline_policy}.
From Figure~\ref{fig:il_rl_w_off_seq1}(a) and Figure~\ref{fig:il_rl_w_off_seq1}(b), it is observed that RL-based online policy does not achieve similar performance as Oracle policy most of the time.
For example, the accuracy of choosing the frequency of big cores ($f_B$) and frequency of little cores ($f_L$) are 100\% only 16\% of the entire execution with RL-based online learned policy. 
Moreover, we observe that the online RL policy could not converge to optimal configuration even after the sequence is run 14 times back to back.}
In \textit{Sequence 2}, we run MotionEstimation, BlackScholes-4T, AES, BlackScholes-2T, Kmeans, Spectral and PCA applications one after another. 
Similar to the previous example, our initial policy converges to the optimal policy only 1.2 seconds, as shown in Figure~\ref{fig:il_rl_w_off_seq2}(a) and Figure~\ref{fig:il_rl_w_off_seq2}(b).
\jrnAdd{RL-based online training is unable to converge as fast as online IL policy for this application sequence also.}
For both sequences, the initial offline IL model is retrained once before converging.
\jrnAdd{However, the offline RL model is not able to obtain high accuracy ever after retraining five times.
We also compare the energy consumption of the applications upon applying the RL policy and the proposed online IL policy.
Figure~\ref{fig:energy_set_1_w_off} and Figure~\ref{fig:energy_set_2_w_off} show this energy comparison with respect to the Oracle for application sequence 1 and application sequence 2 respectively.
For both sequences, we observe that the energy consumption with the proposed online IL policy is always within 2\% of the energy consumption with Oracle policy.
However, the energy consumption due to the RL policy is 10\%--40\% higher than the energy consumption with the Oracle policy.}
This demonstrates the efficiency of our proposed online IL technique to adapt to new unseen applications.

\begin{figure}[t]
	\centering
	\includegraphics[width=1\linewidth]{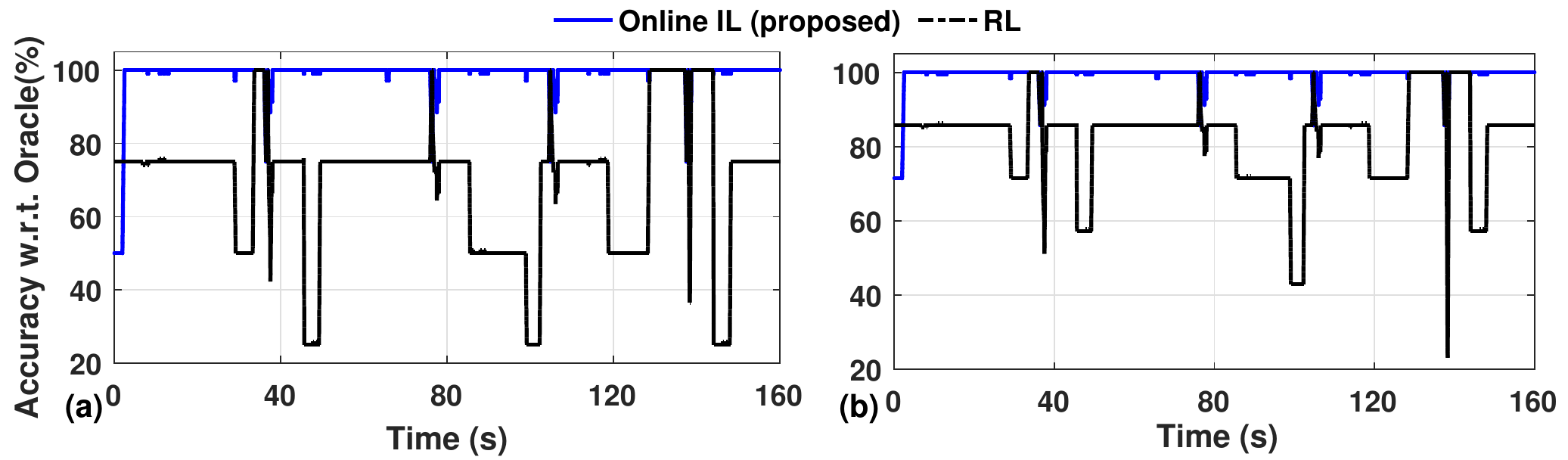}
	\caption{Comparison between proposed online IL and reinforcement learning (RL) with application sequence 1 for (a) frequency of big cores ($f_B$), (b) frequency of little cores ($f_L$).  In this experiment, no pre-trained offline model is used.}
	\label{fig:il_rl_wo_off_seq1}
\end{figure}
\begin{figure}[t]
	\centering
	\includegraphics[width=1\linewidth]{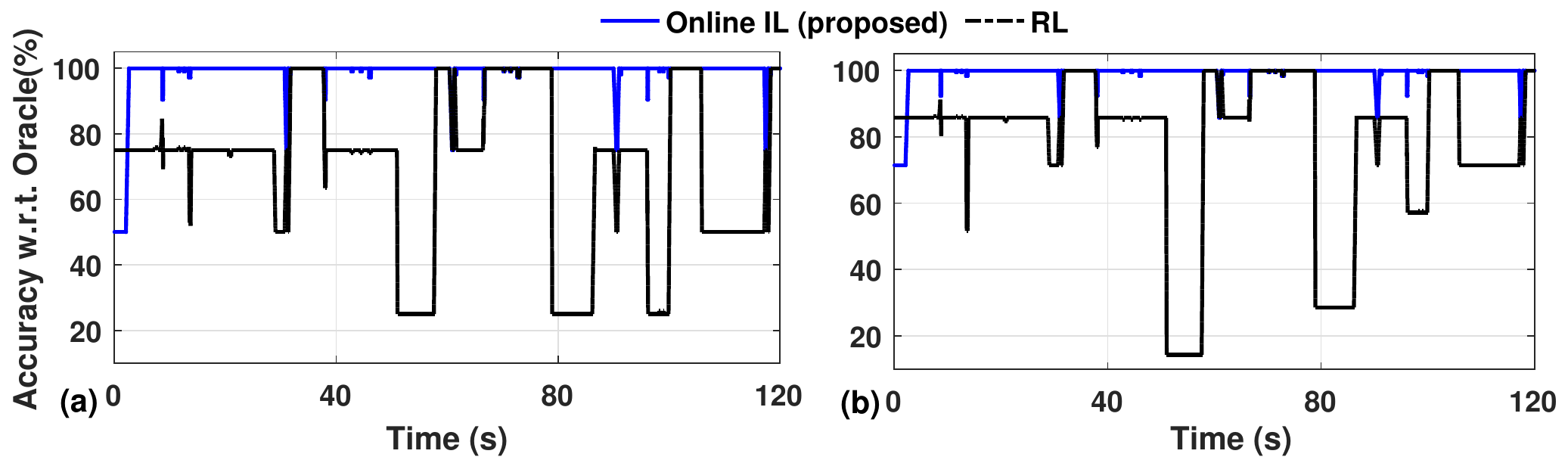}
	\caption{Comparison between proposed online IL and reinforcement learning (RL) with application sequence 2 for (a) frequency of big cores ($f_B$), (b) frequency of little cores ($f_L$). In this experiment, no pre-trained offline model is used.}
	\label{fig:il_rl_wo_off_seq2}
\end{figure}
\begin{figure*}[t]
    \begin{minipage}{0.48\textwidth}
	\centering
	\resizebox{1.0\columnwidth}{!}{\includegraphics[width=0.8\textwidth]{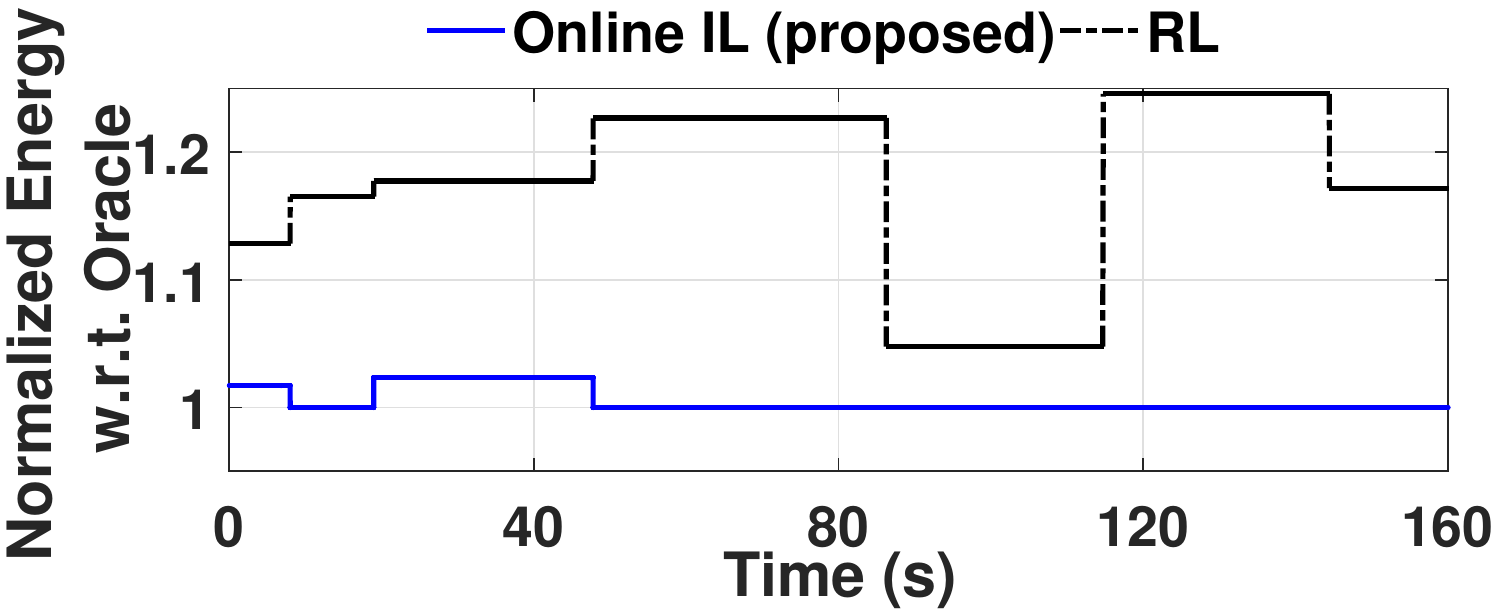}}
	\vspace{-3mm}
	\caption{Energy comparison with application sequence 1 without offline model.}
	\label{fig:energy_set_1_wo_off}
	\end{minipage} \hfill
    \begin{minipage}{0.48\textwidth}
	\centering
	\resizebox{1.0\columnwidth}{!}{\includegraphics[width=0.8\textwidth]{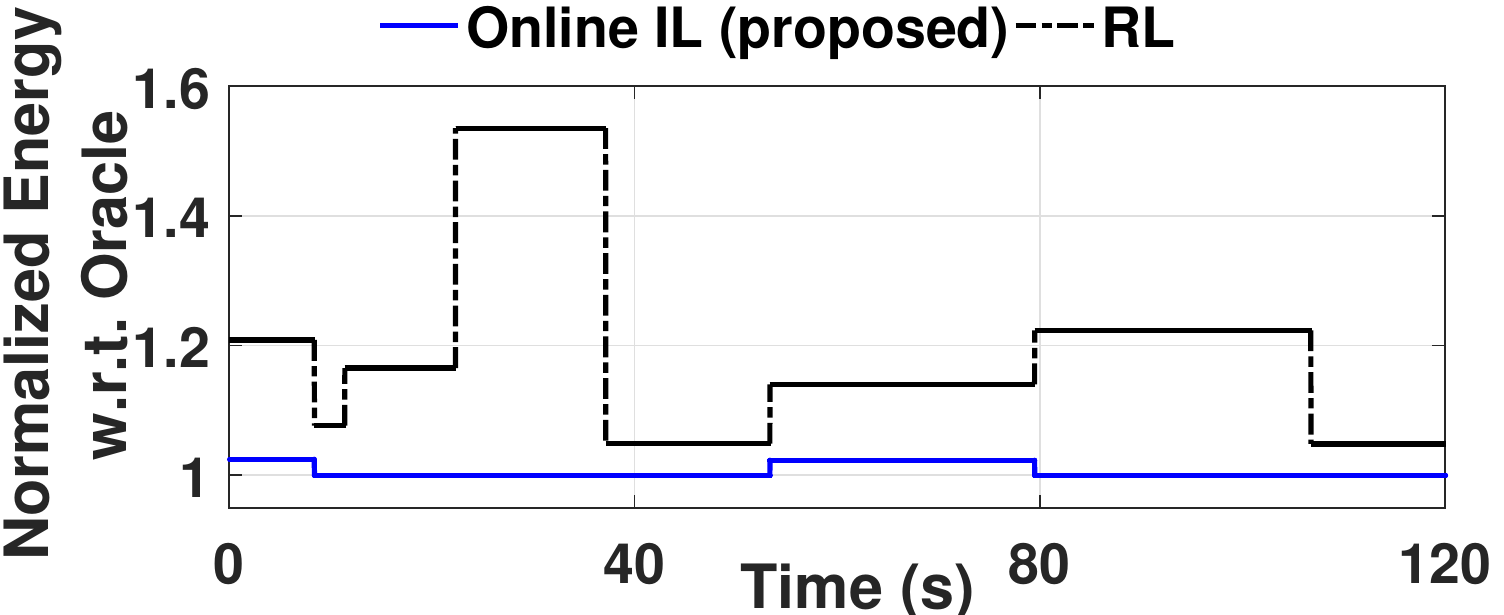}}
	\vspace{-3mm}
	\caption{Energy comparison with application sequence 2 without offline model.}
	\label{fig:energy_set_2_wo_off}
	\end{minipage}
\end{figure*}
\noindent \textbf{Evaluation of the Proposed Approach without Offline Learning:}
We also evaluate the proposed online learning framework when a pre-trained offline model is not available.
In this case, we start with a neural network model initialized with random weights.
\rev{At runtime, with incoming applications, the model is updated using the proposed approach.}
Figure~\ref{fig:il_rl_wo_off_seq1}(a) and Figure~\ref{fig:il_rl_wo_off_seq1}(b) show how fast the online policy converges when the pre-trained offline model is not provided for frequency of big cores ($f_B$) and frequency of little cores ($f_L$) respectively.
In this case also the online IL policy converges within 1.5 second of execution which is around 5\% of the sequence.
Similar to the results with offline policy, policy learned through RL does not obtain high accuracy when compared to the proposed online IL methodology.
Figure~\ref{fig:il_rl_wo_off_seq2}(a) and Figure~\ref{fig:il_rl_wo_off_seq2}(b) show the convergence result with application sequence 2.
In this case also the proposed online IL policy converges faster than RL policy.
The RL policy is unable to achieve optimal performance even after the application sequence is run 5 times.
Figure~\ref{fig:energy_set_1_wo_off} and Figure~\ref{fig:energy_set_2_wo_off} show the energy comparison of the application sequences with RL and proposed online IL method.
In both cases, online IL achieves similar energy consumption as the reference Oracle policy. In contrast, energy consumption with RL policy is always $1.1\times$--$1.6\times$ higher than the energy consumption by the reference Oracle policy.
\begin{figure}[t]
	\centering
	\includegraphics[width=0.8\linewidth]{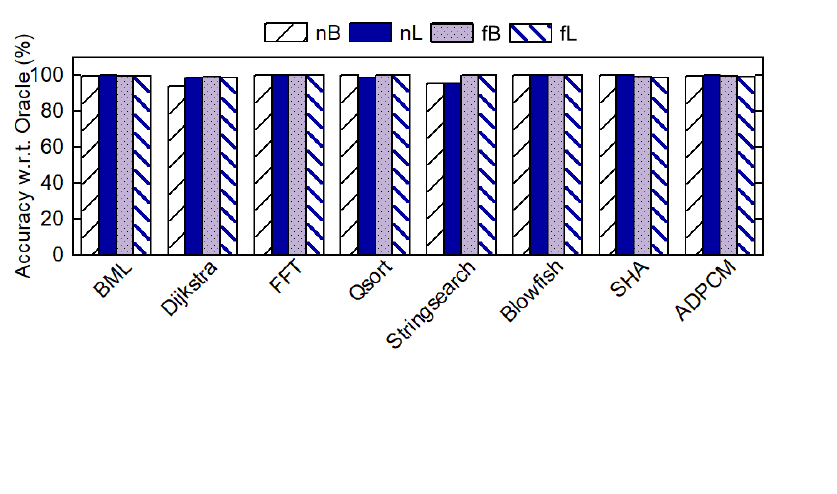}
	\vspace{-23mm}
	\caption{Performance of online learned policy on initial applications.}
	\label{fig:forget_comp}
\end{figure}

\noindent \jrnAdd{\textbf{Performance of Online Learned Policy on Initial Applications:}
In online learning forgetting offline learned policy is a common issue~\cite{goodfellow2013empirical}.
To show that the proposed online IL methodology does not suffer from forgetting, we apply the online learned policy to the applications with which the offline training was performed.
Figure~\ref{fig:forget_comp} shows the accuracy of choosing action knobs for Mibench applications with the policy obtained after online learning is performed on the applications from Cortex and Parsec benchmark suites.
It can be seen that the accuracy is close to $100\%$ for all applications and for all control knobs. This shows that the policy obtained after performing several rounds of online updates does not suffer from unwanted catastrophic forgetting.}

\subsection{Convergence for Single Application}
\begin{figure}[h]
	\centering
	\includegraphics[width=1\linewidth]{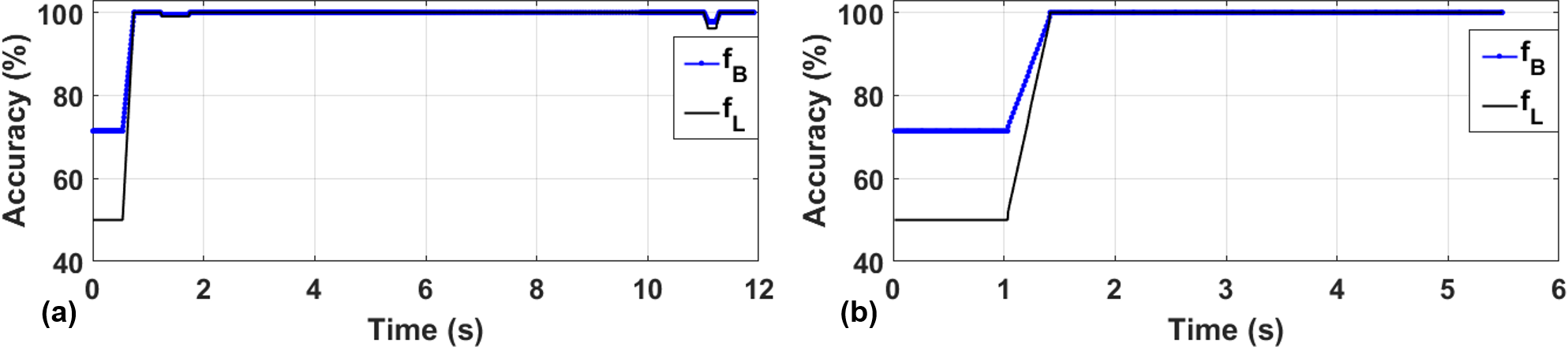}
	\caption{Convergence of $f_B$ and $f_L$ for (a) Spectral and (b) Motion Estimation application.}
	\label{fig:spectral_motionest_time}
\end{figure}

Figure~\ref{fig:accur_all}(c) and Figure~\ref{fig:accur_all}(d) show that 
Kmeans, Spectral, and MotionEstimation applications have large improvements when we perform an online update of the policy.
Therefore, we execute these applications standalone to measure the time required for each of the applications to converge to the optimal Oracle policy.
From Figure~\ref{fig:spectral_motionest_time}(a), we see that to converge to optimal policy Spectral requires 0.7 seconds which is 6\% of the total application execution time.
Also, we observe that within this time our online learning methodology aggregated 100 new samples which are then used to update the policy once.
Similarly, Figure~\ref{fig:spectral_motionest_time}(b) shows that MotionEstimation requires 1.4 seconds to converge.
In this case, 149 new samples are collected as training data to retrain the policy twice.
Finally, we note that each online update of the policy incurs 2 ms of execution time. 
Since this is less than 0.5\% of the time interval between successive updates, 
the runtime overhead is small.

\begin{figure}[t]
	\centering
	\includegraphics[width=0.6\linewidth]{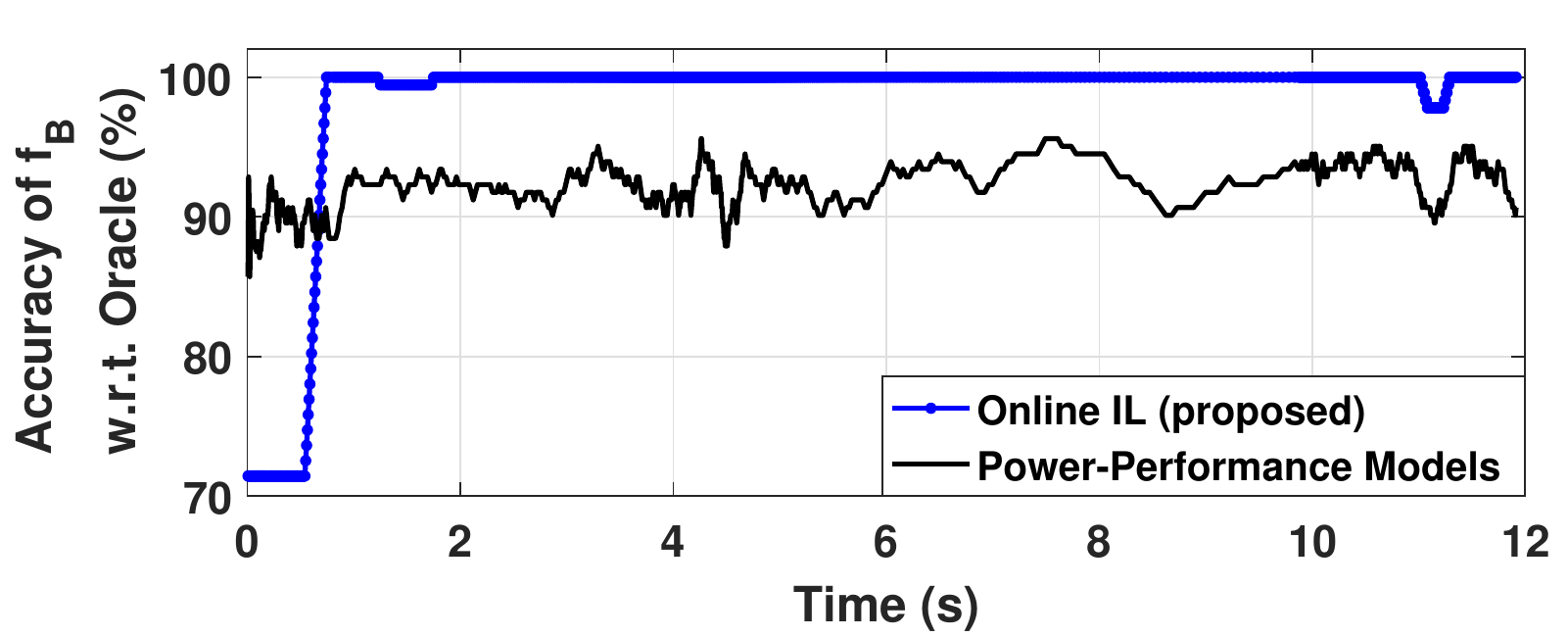}
	\caption{\rev{Analysis of a Policy using only Power-Performance Models.}}
	\label{fig:spectral_pnp_only}
\end{figure}

\noindent\rev{\textbf{Analysis of a Policy using only Power-Performance Models:} We perform another experiment to specifically highlight the contribution of online-IL to provide an optimal policy. In this experiment, we do not use online-IL to estimate the configuration. Instead, we only use the power-performance models to estimate the configuration. Specifically, in the current epoch, we search around the configuration which was found optimal through the search performed in the previous epoch (same as in Algorithm~\ref{alg:oracle_algorithm}). Figure~\ref{fig:spectral_pnp_only} shows the accuracy comparison for frequency of big cores ($f_B$) between proposed online-IL technique and the case where we use only power-performance models. This experiment is performed for the Spectral application. We observe that, if only power-performance models are used, the initial accuracy is more than the proposed method. This happens because, initially, the IL policy is not suitable for the application. However, after learning on a sufficient number of samples, the IL policy reaches to the highest accuracy. This IL policy uses an online Oracle which is obtained through a local search around the current configuration. We note that, if there is no learned policy, the local search guided only by the power-performance models would not be effective. The online learned policy provides a good starting configuration to bootstrap the search process. In other words, only a small amount of search is sufficient to reach the optimal configuration if we perform local search around the configuration suggested by the learned policy. On the other hand, if only power-performance models are used to take the control decision, then local search will require evaluating a significantly large number of configurations (as much as 3$\times$ than online-IL) to construct an equally good online Oracle policy, which is not practically feasible.  Therefore, if we perform local search guided only by the power-performance models, the resulting policy never converges to the optimal as shown in Figure~\ref{fig:spectral_pnp_only}. Since local search requires a significantly large number of evaluations of configurations using power-performance models, it is not practical to use only power-performance models for online control. The combination of learned policy for getting good initial configuration and a small amount of local search guided by power-performance models to generate better configurations (supervision to improve policy) makes our overall approach effective. Once we converge to a near-optimal policy, we can apply the policy with very little search, but the baseline solution (search guided by power-performance models) will still require a lot of search and is impractical. }



%

\begin{figure}[t]
	\centering
	\includegraphics[width=0.8\linewidth]{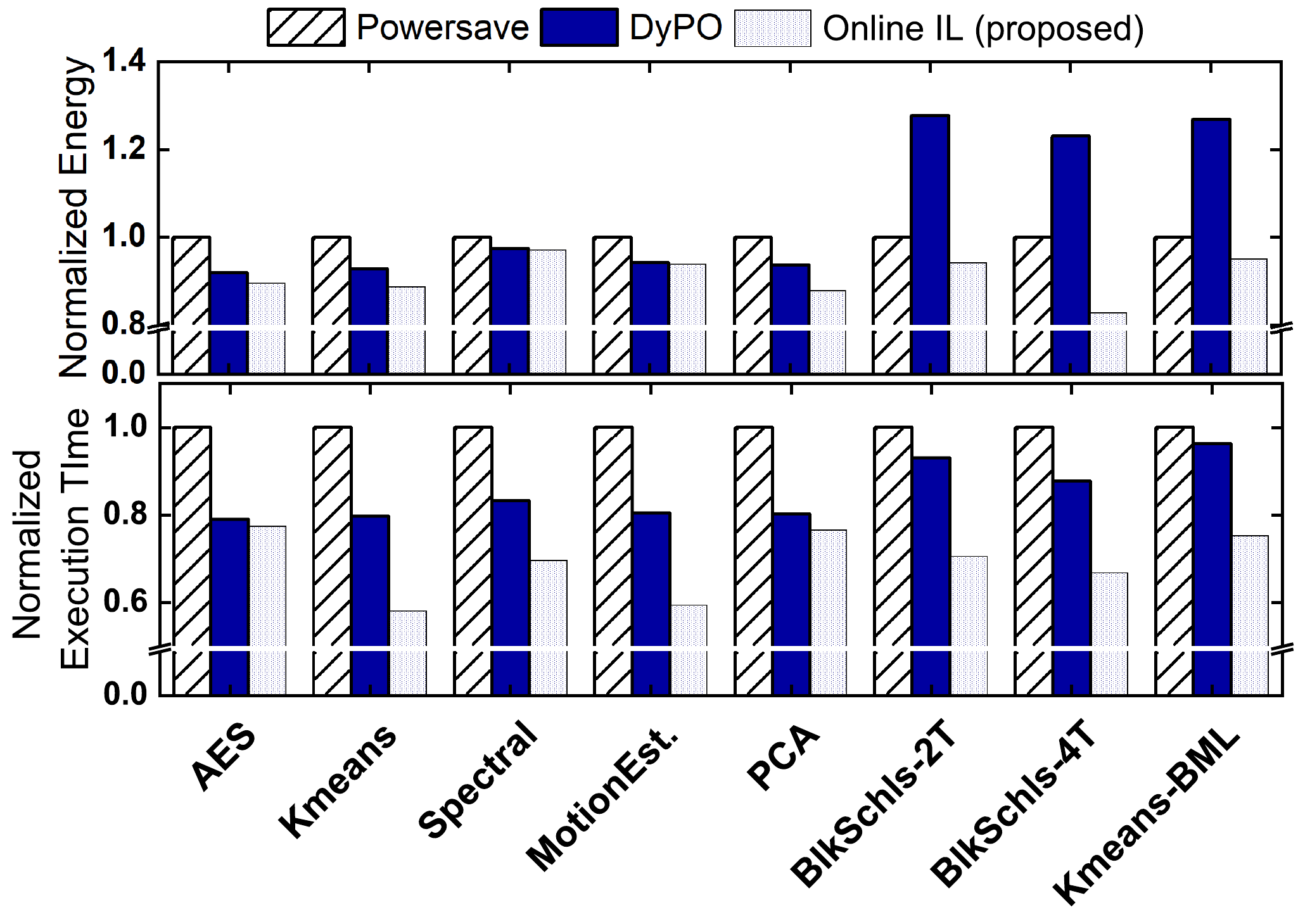}
	\caption{\rev{Comparison of energy consumption and execution time with respect to the DyPO~\cite{gupta2017dypo} and Powersave governor.}}
	\label{fig:pwrsave_gov}
\end{figure}

\jrnAdd{\subsection{Comparison with the state-of-the-art Power Management Techniques}}
Powersave governor is commonly used in mobile platforms to operate in low power mode.
Therefore, we compare the energy consumption of individual applications 
when we use the proposed online-IL approach and the Powersave governor. 
We note that these comparisons include energy and execution time overheads 
since the evaluations are based on hardware measurements. 
Similar to the previous sections, we start with an offline policy trained with applications from Mibench suite.
Then, we use the proposed technique to update the initial policy as we run applications from Cortex and PARSEC suites.

Figure~\ref{fig:pwrsave_gov} shows the energy consumption and execution time comparisons.
We normalize the energy and execution time of all applications with respect to the Powersave governor since absolute numbers vary widely across different applications.
We observe that the proposed online learning methodology achieves \textit{both lower energy and execution time} compared to the Powersave governor for all applications.
More specifically, we achieve on average 10\% lower energy consumption and
24\% faster execution time.
In particular, we improve the execution time for Kmeans by 42\% 
while also reducing the energy consumption.
%

The proposed policy achieves 48.4\% less energy consumption on average when compared to the interactive governor.
We omit the detailed comparison with the interactive governor since it targets performance optimization.
We also compared the proposed online learning methodology with a recent offline learning technique~\cite{gupta2017dypo}. The DyPO approach first finds the Pareto-optimal configurations for energy optimal operation and then chooses a configuration among them at runtime. However, the DyPO approach does not perform any online learning.
We observe that our proposed methodology on average achieves 6.39\% improvement in energy consumption and 15.24\% improvement in execution time for unseen applications when compared to the DyPO approach.

\rev{Furthermore, to validate the robustness of the proposed online IL technique, we apply the technique while executing Kmeans and BML applications concurrently (Kmeans-BML). We choose Kmeans as the foreground application and BML as the background application. We observe that the proposed online IL technique results in 5\% less energy consumption than powersave governor for Kmeans-BML. Moreover, the execution time of Kmeans-BML with proposed online IL technique is 25\% and 22\% less than the powersave governor and DyPO respectively. This experiment proves that the proposed online IL technique can efficiently manage the platform while executing multiple applications concurrently.}

In summary, the proposed online imitation learning methodology efficiently adapts to unseen applications and provides significant improvements over state-of-the-art methods with minimal overhead.

\begin{table}[t]
\centering
\caption{Runtime and implementation overhead summary.} \label{tab:overhead}
\begin{tabular}{|c|c|c|c|c|}
\hline
\begin{tabular}[c]{@{}c@{}}Metric\end{tabular}                  & Module                                                        & Overhead     & \begin{tabular}[c]{@{}c@{}}Time between\\ successive calls\end{tabular} & \begin{tabular}[c]{@{}c@{}}\% Over-\\ head\end{tabular} \\ \hline
\multirow{3}{*}{\begin{tabular}[c]{@{}c@{}}Exe. \\ time\end{tabular}} & RLS                                                           & 6 $\mu$s    & 10-20 ms                                                                & 0.06-0.12                                                      \\ \cline{2-5} 
                                                                           & \begin{tabular}[c]{@{}c@{}}Local \\ search\end{tabular}       & 2.42~$\mu$s & 10-20 ms                                                                & 0.02-0.05                                                     \\ \cline{2-5} 
                                                                           & Backprop.                                                     & 2 ms         & 0.5-1.5 s                                                               & 0.4-1.0                                                         \\ \hline
\end{tabular}
\end{table}

\vspace{1.0ex}

\subsection{Implementation Overhead Analysis}

\label{sec:overhead}
We conclude our experimental evaluation by summarizing the runtime and storage overhead of our approach measured on Odroid XU3 board.  The online power-performance models are updated in each epoch, which ranges between 10--20ms. Since the model updates take 6$\mu s$, the runtime overhead is only 0.06-0.12\%, as summarized in Table~\ref{tab:overhead}. 
Online policy update via backpropagation takes longer (2ms) than the other 
operations, as expected. 
However, we note that it runs every 100 epochs, i.e., in the order of seconds. Therefore, its runtime overhead is less than 1\% as shown in Table~\ref{tab:overhead}. 
Finally, the storage overhead of the buffer to store new training examples is only 18KB. 
In summary, the proposed approach incurs minimal runtime and storage overheads.

\noindent\rev{\textbf{Overhead Comparison with RL:} We start the overhead comparison with the overhead of retraining neural network with backpropagation algorithm. For each backpropagation step we use 100 training samples for both online-IL and RL algorithms. Furthermore, the network structure used in both the algorithm is similar. Therefore, the execution time overhead for both algorithms for a single backpropagation step is same (2 ms). However, RL needs more backpropagation steps than the proposed online-IL to learn a policy that performs as good as online-IL. As a result, the cumulative overhead of RL in terms of both execution time and training samples required is much higher than online-IL. For an example, to obtain the optimal policy for Spectral application, online-IL performs backpropagation step only once (total 2 ms overhead). But, even after RL performs 8 backpropagation steps (total 16 ms overhead), it is not able to learn the optimal policy for Spectral. Overall, for Spectral application, the performance overhead incurred by RL is twice of the overhead incurred by IL. This further proves the efficiency of proposed online-IL approach over RL.}



\section{Conclusions and Future Work}

Designing an optimal runtime power management policy that achieves fast response time and minimum energy consumption is challenging due to the large space of available control knobs and diversity of applications. Policies constructed at design-time may fail to deliver optimal execution for applications encountered at runtime. This paper presented a novel online imitation learning methodology to efficiently learn the optimal policy for new applications at runtime. Experimental evaluation on Odroid XU3 platform shows that we achieve both more than 10\% lower energy consumption 
and 24\% faster execution time with respect to the Powersave governor.
\rev{The proposed approach is also applicable to other sequential decision-making problems in mobile systems, such as, like prefetcher, cache replacement, branch prediction. However, extending the proposed IL framework to these domains is beyond the scope of this paper. Therefore, it is left as future work.}

\bibliographystyle{ACM-Reference-Format}
\bibliography{IL_TODAES.bbl}


\begin{thebibliography}{48}


\ifx \showCODEN    \undefined \def \showCODEN     #1{\unskip}     \fi
\ifx \showDOI      \undefined \def \showDOI       #1{#1}\fi
\ifx \showISBNx    \undefined \def \showISBNx     #1{\unskip}     \fi
\ifx \showISBNxiii \undefined \def \showISBNxiii  #1{\unskip}     \fi
\ifx \showISSN     \undefined \def \showISSN      #1{\unskip}     \fi
\ifx \showLCCN     \undefined \def \showLCCN      #1{\unskip}     \fi
\ifx \shownote     \undefined \def \shownote      #1{#1}          \fi
\ifx \showarticletitle \undefined \def \showarticletitle #1{#1}   \fi
\ifx \showURL      \undefined \def \showURL       {\relax}        \fi
\providecommand\bibfield[2]{#2}
\providecommand\bibinfo[2]{#2}
\providecommand\natexlab[1]{#1}
\providecommand\showeprint[2][]{arXiv:#2}

\bibitem[\protect\citeauthoryear{Aalsaud et~al\mbox{.}}{Aalsaud
  et~al\mbox{.}}{2016}]%
        {aalsaud2016power}
\bibfield{author}{\bibinfo{person}{Ali Aalsaud} {et~al\mbox{.}}}
  \bibinfo{year}{2016}\natexlab{}.
\newblock \showarticletitle{{Power--Aware Performance Adaptation of Concurrent
  Applications In Heterogeneous Many-Core Systems}}. In
  \bibinfo{booktitle}{\emph{Proc. Int. Symp. on Low Power Electronics and
  Design}}. \bibinfo{pages}{368--373}.
\newblock


\bibitem[\protect\citeauthoryear{Bellosa}{Bellosa}{2000}]%
        {bellosa2000benefits}
\bibfield{author}{\bibinfo{person}{Frank Bellosa}.}
  \bibinfo{year}{2000}\natexlab{}.
\newblock \showarticletitle{{The Benefits of Event: Driven Energy Accounting in
  Power-Sensitive Systems}}. In \bibinfo{booktitle}{\emph{Proceedings of the
  9th workshop on ACM SIGOPS European workshop: beyond the PC: new challenges
  for the operating system}}. \bibinfo{pages}{37--42}.
\newblock


\bibitem[\protect\citeauthoryear{Bhat, Mandal, Gupta, and Ogras}{Bhat
  et~al\mbox{.}}{2018}]%
        {bhat2018online}
\bibfield{author}{\bibinfo{person}{Ganapati Bhat}, \bibinfo{person}{Sumit~K
  Mandal}, \bibinfo{person}{Ujjwal Gupta}, {and} \bibinfo{person}{Umit~Y
  Ogras}.} \bibinfo{year}{2018}\natexlab{}.
\newblock \showarticletitle{{Online learning for adaptive optimization of
  heterogeneous SoCs}}. In \bibinfo{booktitle}{\emph{Proceedings of the
  International Conference on Computer-Aided Design}}. ACM,
  \bibinfo{pages}{61}.
\newblock


\bibitem[\protect\citeauthoryear{Bhat, Singla, Unver, and Ogras}{Bhat
  et~al\mbox{.}}{2017}]%
        {bhat2017algorithmic}
\bibfield{author}{\bibinfo{person}{Ganapati Bhat}, \bibinfo{person}{Gaurav
  Singla}, \bibinfo{person}{Ali~K Unver}, {and} \bibinfo{person}{Umit~Y
  Ogras}.} \bibinfo{year}{2017}\natexlab{}.
\newblock \showarticletitle{{Algorithmic Optimization of Thermal and Power
  Management for Heterogeneous Mobile Platforms}}.
\newblock \bibinfo{journal}{\emph{IEEE Transactions on Very Large Scale
  Integration (VLSI) Systems}} \bibinfo{volume}{26}, \bibinfo{number}{3}
  (\bibinfo{year}{2017}), \bibinfo{pages}{544--557}.
\newblock


\bibitem[\protect\citeauthoryear{Bienia et~al\mbox{.}}{Bienia
  et~al\mbox{.}}{2008}]%
        {bienia2008parsec}
\bibfield{author}{\bibinfo{person}{Christian Bienia} {et~al\mbox{.}}}
  \bibinfo{year}{2008}\natexlab{}.
\newblock \showarticletitle{{The PARSEC Benchmark Suite: Characterization and
  Architectural Implications}}. In \bibinfo{booktitle}{\emph{Proc. Int. Conf.
  on Parallel Arch. and Compilation Techniques}}. \bibinfo{pages}{72--81}.
\newblock


\bibitem[\protect\citeauthoryear{Brooks, Dick, Joseph, and Shang}{Brooks
  et~al\mbox{.}}{2007}]%
        {brooks2007power}
\bibfield{author}{\bibinfo{person}{David Brooks}, \bibinfo{person}{Robert~P
  Dick}, \bibinfo{person}{Russ Joseph}, {and} \bibinfo{person}{Li Shang}.}
  \bibinfo{year}{2007}\natexlab{}.
\newblock \showarticletitle{{Power, Thermal, and Reliability Modeling in
  Nanometer-Scale Microprocessors}}.
\newblock \bibinfo{journal}{\emph{IEEE Micro}} \bibinfo{volume}{27},
  \bibinfo{number}{3} (\bibinfo{year}{2007}), \bibinfo{pages}{49--62}.
\newblock


\bibitem[\protect\citeauthoryear{Chen and Marculescu}{Chen and
  Marculescu}{2015}]%
        {chen2015distributed}
\bibfield{author}{\bibinfo{person}{Zhuo Chen} {and} \bibinfo{person}{Diana
  Marculescu}.} \bibinfo{year}{2015}\natexlab{}.
\newblock \showarticletitle{{Distributed Reinforcement Learning For Power
  Limited Many-Core System Performance Optimization}}. In
  \bibinfo{booktitle}{\emph{Proc. of the DATE}}. \bibinfo{pages}{1521--1526}.
\newblock


\bibitem[\protect\citeauthoryear{Cochran et~al\mbox{.}}{Cochran
  et~al\mbox{.}}{[n. d.]}]%
        {cochran2011pack}
\bibfield{author}{\bibinfo{person}{Ryan Cochran} {et~al\mbox{.}}}
  \bibinfo{year}{[n. d.]}\natexlab{}.
\newblock \showarticletitle{{Pack \& Cap: adaptive DVFS and thread packing
  under power caps}}. In \bibinfo{booktitle}{\emph{2011 44th Annual IEEE/ACM
  Intl. Symp. on Microarchitecture (MICRO)}}. \bibinfo{pages}{175--185}.
\newblock


\bibitem[\protect\citeauthoryear{Dhiman et~al\mbox{.}}{Dhiman
  et~al\mbox{.}}{2009}]%
        {dhiman2009system}
\bibfield{author}{\bibinfo{person}{Gaurav Dhiman} {et~al\mbox{.}}}
  \bibinfo{year}{2009}\natexlab{}.
\newblock \showarticletitle{{System-level power management using online
  learning}}.
\newblock \bibinfo{journal}{\emph{IEEE Trans. on CAD}} \bibinfo{volume}{28},
  \bibinfo{number}{5} (\bibinfo{year}{2009}), \bibinfo{pages}{676--689}.
\newblock


\bibitem[\protect\citeauthoryear{Dietrich and Chakraborty}{Dietrich and
  Chakraborty}{2014}]%
        {dietrich2014lightweight}
\bibfield{author}{\bibinfo{person}{Benedikt Dietrich} {and}
  \bibinfo{person}{Samarjit Chakraborty}.} \bibinfo{year}{2014}\natexlab{}.
\newblock \showarticletitle{{Lightweight Graphics Instrumentation for Game
  State-Specific Power Management in Android}}.
\newblock \bibinfo{journal}{\emph{Multimedia Systems}} \bibinfo{volume}{20},
  \bibinfo{number}{5} (\bibinfo{year}{2014}), \bibinfo{pages}{563--578}.
\newblock


\bibitem[\protect\citeauthoryear{Dietrich, Nunna, Goswami, Chakraborty, and
  Gries}{Dietrich et~al\mbox{.}}{2010}]%
        {dietrich2010lms}
\bibfield{author}{\bibinfo{person}{Benedikt Dietrich}, \bibinfo{person}{Swaroop
  Nunna}, \bibinfo{person}{Dip Goswami}, \bibinfo{person}{Samarjit
  Chakraborty}, {and} \bibinfo{person}{Matthias Gries}.}
  \bibinfo{year}{2010}\natexlab{}.
\newblock \showarticletitle{{LMS-based Low-Complexity Game Workload Prediction
  for DVFS}}. In \bibinfo{booktitle}{\emph{2010 IEEE International Conference
  on Computer Design}}. \bibinfo{pages}{417--424}.
\newblock


\bibitem[\protect\citeauthoryear{Ge and Qiu}{Ge and Qiu}{2011}]%
        {ge2011dynamic}
\bibfield{author}{\bibinfo{person}{Yang Ge} {and} \bibinfo{person}{Qinru Qiu}.}
  \bibinfo{year}{2011}\natexlab{}.
\newblock \showarticletitle{{Dynamic Thermal Management for Multimedia
  Applications using Machine Learning}}. In
  \bibinfo{booktitle}{\emph{Proceedings of the 48th Design Automation
  Conference}}. \bibinfo{pages}{95--100}.
\newblock


\bibitem[\protect\citeauthoryear{Goodfellow et~al\mbox{.}}{Goodfellow
  et~al\mbox{.}}{2013}]%
        {goodfellow2013empirical}
\bibfield{author}{\bibinfo{person}{Ian~J Goodfellow} {et~al\mbox{.}}}
  \bibinfo{year}{2013}\natexlab{}.
\newblock \showarticletitle{{An Empirical Investigation of Catastrophic
  Forgetting in Gradient-based Neural Networks}}.
\newblock \bibinfo{journal}{\emph{arXiv preprint arXiv:1312.6211}}
  (\bibinfo{year}{2013}).
\newblock


\bibitem[\protect\citeauthoryear{Gupta et~al\mbox{.}}{Gupta
  et~al\mbox{.}}{2017}]%
        {gupta2017dypo}
\bibfield{author}{\bibinfo{person}{Ujjwal Gupta} {et~al\mbox{.}}}
  \bibinfo{year}{2017}\natexlab{}.
\newblock \showarticletitle{{DyPO: Dynamic Pareto-Optimal Configuration
  Selection For Heterogeneous MpSoCs}}.
\newblock \bibinfo{journal}{\emph{ACM Trans. Embedd. Comput. Syst.}}
  \bibinfo{volume}{16}, \bibinfo{number}{5s} (\bibinfo{year}{2017}),
  \bibinfo{pages}{123}.
\newblock


\bibitem[\protect\citeauthoryear{Gupta et~al\mbox{.}}{Gupta
  et~al\mbox{.}}{2019}]%
        {gupta2019deep}
\bibfield{author}{\bibinfo{person}{Ujjwal Gupta} {et~al\mbox{.}}}
  \bibinfo{year}{2019}\natexlab{}.
\newblock \showarticletitle{{A Deep Q-Learning Approach for Dynamic Management
  of Heterogeneous Processors}}.
\newblock \bibinfo{journal}{\emph{IEEE Computer Architecture Letters}}
  (\bibinfo{year}{2019}).
\newblock


\bibitem[\protect\citeauthoryear{Gupta, Babu, Ayoub, Kishinevsky, Paterna,
  Gumussoy, and Ogras}{Gupta et~al\mbox{.}}{2018b}]%
        {gupta2018online}
\bibfield{author}{\bibinfo{person}{Ujjwal Gupta}, \bibinfo{person}{Manoj Babu},
  \bibinfo{person}{Raid Ayoub}, \bibinfo{person}{Michael Kishinevsky},
  \bibinfo{person}{Francesco Paterna}, \bibinfo{person}{Suat Gumussoy}, {and}
  \bibinfo{person}{Umit~Y Ogras}.} \bibinfo{year}{2018}\natexlab{b}.
\newblock \showarticletitle{{An Online Learning Methodology for Performance
  Modeling of Graphics Processors}}.
\newblock \bibinfo{journal}{\emph{IEEE Trans. Comput.}} \bibinfo{volume}{67},
  \bibinfo{number}{12} (\bibinfo{year}{2018}), \bibinfo{pages}{1677--1691}.
\newblock


\bibitem[\protect\citeauthoryear{Gupta, Babu, Ayoub, Kishinevsky, Paterna, and
  Ogras}{Gupta et~al\mbox{.}}{2018a}]%
        {gupta2018staff}
\bibfield{author}{\bibinfo{person}{Ujjwal Gupta}, \bibinfo{person}{Manoj Babu},
  \bibinfo{person}{Raid Ayoub}, \bibinfo{person}{Michael Kishinevsky},
  \bibinfo{person}{Francesco Paterna}, {and} \bibinfo{person}{Umit~Y Ogras}.}
  \bibinfo{year}{2018}\natexlab{a}.
\newblock \showarticletitle{{STAFF: Online Learning with Stabilized Adaptive
  Forgetting Factor and Feature Selection Algorithm}}. In
  \bibinfo{booktitle}{\emph{Proceedings of the 55th Annual Design Automation
  Conference}}. \bibinfo{pages}{1--6}.
\newblock


\bibitem[\protect\citeauthoryear{Guthaus et~al\mbox{.}}{Guthaus
  et~al\mbox{.}}{2001}]%
        {guthaus2001mibench}
\bibfield{author}{\bibinfo{person}{Matthew~R Guthaus} {et~al\mbox{.}}}
  \bibinfo{year}{2001}\natexlab{}.
\newblock \showarticletitle{{Mibench: A Free, Commercially Representative
  Embedded Benchmark Suite}}. In \bibinfo{booktitle}{\emph{Proc. of the Int.
  Workshop on Workload Characterization}}. \bibinfo{pages}{3--14}.
\newblock


\bibitem[\protect\citeauthoryear{{Hardkernel}}{{Hardkernel}}{2014}]%
        {ODROID_Platforms}
\bibfield{author}{\bibinfo{person}{{Hardkernel}}.}
  \bibinfo{year}{2014}\natexlab{}.
\newblock \bibinfo{title}{{ODROID-XU3}}.
\newblock
  \bibinfo{howpublished}{\url{https://wiki.odroid.com/old_product/odroid-xu3/odroid-xu3}
  Accessed 24 Nov. 2018}.
\newblock


\bibitem[\protect\citeauthoryear{Hecht-Nielsen}{Hecht-Nielsen}{1992}]%
        {hecht1992theory}
\bibfield{author}{\bibinfo{person}{Robert Hecht-Nielsen}.}
  \bibinfo{year}{1992}\natexlab{}.
\newblock \showarticletitle{{Theory of the backpropagation neural network}}.
\newblock In \bibinfo{booktitle}{\emph{Neural networks for perception}}.
  \bibinfo{publisher}{Elsevier}, \bibinfo{pages}{65--93}.
\newblock


\bibitem[\protect\citeauthoryear{Kadjo, Ayoub, Kishinevsky, and Gratz}{Kadjo
  et~al\mbox{.}}{2015}]%
        {kadjo2015control}
\bibfield{author}{\bibinfo{person}{David Kadjo}, \bibinfo{person}{Raid Ayoub},
  \bibinfo{person}{Michael Kishinevsky}, {and} \bibinfo{person}{Paul~V Gratz}.}
  \bibinfo{year}{2015}\natexlab{}.
\newblock \showarticletitle{{A Control-Theoretic Approach for Energy Efficient
  CPU-GPU Subsystem in Mobile Platforms}}. In
  \bibinfo{booktitle}{\emph{Proceedings of the 52nd Annual Design Automation
  Conference}}. ACM, \bibinfo{pages}{62}.
\newblock


\bibitem[\protect\citeauthoryear{Kim et~al\mbox{.}}{Kim et~al\mbox{.}}{2017}]%
        {kim2017imitation}
\bibfield{author}{\bibinfo{person}{Ryan~Gary Kim} {et~al\mbox{.}}}
  \bibinfo{year}{2017}\natexlab{}.
\newblock \showarticletitle{{Imitation Learning For Dynamic VFI Control In
  Large-Scale Manycore Systems}}.
\newblock \bibinfo{journal}{\emph{IEEE Trans. on VLSI Systems}}
  \bibinfo{volume}{25}, \bibinfo{number}{9} (\bibinfo{year}{2017}),
  \bibinfo{pages}{2458--2471}.
\newblock


\bibitem[\protect\citeauthoryear{Lattner and Adve}{Lattner and Adve}{2004}]%
        {lattner2004llvm}
\bibfield{author}{\bibinfo{person}{Chris Lattner} {and} \bibinfo{person}{Vikram
  Adve}.} \bibinfo{year}{2004}\natexlab{}.
\newblock \showarticletitle{{LLVM: A compilation Framework For Lifelong Program
  Analysis \& Transformation}}. In \bibinfo{booktitle}{\emph{Proc. of the Int.
  Symp. on Code Generation and Optimization: Feedback-Directed and Runtime
  Optimization}}. \bibinfo{pages}{75}.
\newblock


\bibitem[\protect\citeauthoryear{Mandal et~al\mbox{.}}{Mandal
  et~al\mbox{.}}{2019}]%
        {mandal2019dynamic}
\bibfield{author}{\bibinfo{person}{Sumit~K Mandal} {et~al\mbox{.}}}
  \bibinfo{year}{2019}\natexlab{}.
\newblock \showarticletitle{{Dynamic Resource Management of Heterogeneous
  Mobile Platforms via Imitation Learning}}.
\newblock \bibinfo{journal}{\emph{IEEE Transactions on Very Large Scale
  Integration (VLSI) Systems}} (\bibinfo{year}{2019}).
\newblock


\bibitem[\protect\citeauthoryear{Martinez and Ipek}{Martinez and Ipek}{2009}]%
        {martinez2009dynamic}
\bibfield{author}{\bibinfo{person}{Jose~F Martinez} {and}
  \bibinfo{person}{Engin Ipek}.} \bibinfo{year}{2009}\natexlab{}.
\newblock \showarticletitle{{Dynamic Multicore Resource Management: A Machine
  Learning Approach}}.
\newblock \bibinfo{journal}{\emph{IEEE Micro}} \bibinfo{volume}{29},
  \bibinfo{number}{5} (\bibinfo{year}{2009}).
\newblock


\bibitem[\protect\citeauthoryear{Mendel}{Mendel}{1995}]%
        {mendel1995lessons}
\bibfield{author}{\bibinfo{person}{Jerry~M Mendel}.}
  \bibinfo{year}{1995}\natexlab{}.
\newblock \bibinfo{booktitle}{\emph{{Lessons in estimation theory for signal
  processing, communications, and control}}}.
\newblock \bibinfo{publisher}{Pearson Education}.
\newblock


\bibitem[\protect\citeauthoryear{Mnih et~al\mbox{.}}{Mnih
  et~al\mbox{.}}{2015}]%
        {mnih2015human}
\bibfield{author}{\bibinfo{person}{Volodymyr Mnih} {et~al\mbox{.}}}
  \bibinfo{year}{2015}\natexlab{}.
\newblock \showarticletitle{{Human-level Control Through Deep Reinforcement
  Learning}}.
\newblock \bibinfo{journal}{\emph{Nature}} \bibinfo{volume}{518},
  \bibinfo{number}{7540} (\bibinfo{year}{2015}), \bibinfo{pages}{529}.
\newblock


\bibitem[\protect\citeauthoryear{Mucci, Browne, Deane, and Ho}{Mucci
  et~al\mbox{.}}{1999}]%
        {mucci1999papi}
\bibfield{author}{\bibinfo{person}{Philip~J Mucci}, \bibinfo{person}{Shirley
  Browne}, \bibinfo{person}{Christine Deane}, {and} \bibinfo{person}{George
  Ho}.} \bibinfo{year}{1999}\natexlab{}.
\newblock \showarticletitle{{PAPI: A Portable Interface To Hardware Performance
  Counters}}. In \bibinfo{booktitle}{\emph{Proc. of the Dept. of defense HPCMP
  Users Group Conf.}}, Vol.~\bibinfo{volume}{710}.
\newblock


\bibitem[\protect\citeauthoryear{Pallipadi and Starikovskiy}{Pallipadi and
  Starikovskiy}{2006}]%
        {pallipadi2006ondemand}
\bibfield{author}{\bibinfo{person}{Venkatesh Pallipadi} {and}
  \bibinfo{person}{Alexey Starikovskiy}.} \bibinfo{year}{2006}\natexlab{}.
\newblock \showarticletitle{{The Ondemand Governor}}. In
  \bibinfo{booktitle}{\emph{Proc. Linux Symp.}}, Vol.~\bibinfo{volume}{2}.
  \bibinfo{pages}{215--230}.
\newblock


\bibitem[\protect\citeauthoryear{Park, Dutt, and Lim}{Park
  et~al\mbox{.}}{2017}]%
        {park2017ml}
\bibfield{author}{\bibinfo{person}{Jurn-Gyu Park}, \bibinfo{person}{Nikil
  Dutt}, {and} \bibinfo{person}{Sung-Soo Lim}.}
  \bibinfo{year}{2017}\natexlab{}.
\newblock \showarticletitle{{ML-Gov: A Machine Learning Enhanced Integrated
  CPU-GPU DVFS Governor For Mobile Gaming}}. In \bibinfo{booktitle}{\emph{Proc.
  Symp. on Embedd. Syst. for Real-Time Multimedia}}. \bibinfo{pages}{12--21}.
\newblock


\bibitem[\protect\citeauthoryear{Pathania et~al\mbox{.}}{Pathania
  et~al\mbox{.}}{2015}]%
        {pathania2015power}
\bibfield{author}{\bibinfo{person}{Anuj Pathania} {et~al\mbox{.}}}
  \bibinfo{year}{2015}\natexlab{}.
\newblock \showarticletitle{{Power-Performance Modelling of Mobile Gaming
  Workloads On Heterogeneous MPSoCs}}. In \bibinfo{booktitle}{\emph{Proc. of
  Design Autom. Conf.}} \bibinfo{pages}{201}.
\newblock


\bibitem[\protect\citeauthoryear{Pathania, Jiao, Prakash, and Mitra}{Pathania
  et~al\mbox{.}}{2014}]%
        {pathania2014integrated}
\bibfield{author}{\bibinfo{person}{Anuj Pathania}, \bibinfo{person}{Qing Jiao},
  \bibinfo{person}{Alok Prakash}, {and} \bibinfo{person}{Tulika Mitra}.}
  \bibinfo{year}{2014}\natexlab{}.
\newblock \showarticletitle{{Integrated CPU-GPU Power Management For 3D Mobile
  Games}}. In \bibinfo{booktitle}{\emph{Design Autom. Conf.}}
  \bibinfo{pages}{1--6}.
\newblock


\bibitem[\protect\citeauthoryear{Ross, Gordon, and Bagnell}{Ross
  et~al\mbox{.}}{2011}]%
        {ross2011reduction}
\bibfield{author}{\bibinfo{person}{St{\'e}phane Ross},
  \bibinfo{person}{Geoffrey Gordon}, {and} \bibinfo{person}{Drew Bagnell}.}
  \bibinfo{year}{2011}\natexlab{}.
\newblock \showarticletitle{{A Reduction of Imitation Learning And Structured
  Prediction To No-Regret Online Learning}}. In \bibinfo{booktitle}{\emph{Proc.
  of the Int. Conf. on Art. Intel. and Stat.}} \bibinfo{pages}{627--635}.
\newblock


\bibitem[\protect\citeauthoryear{Schaal}{Schaal}{1999}]%
        {schaal1999imitation}
\bibfield{author}{\bibinfo{person}{Stefan Schaal}.}
  \bibinfo{year}{1999}\natexlab{}.
\newblock \showarticletitle{{Is Imitation Learning The Route To Humanoid
  Robots?}}
\newblock \bibinfo{journal}{\emph{Trends in cognitive sciences}}
  \bibinfo{volume}{3}, \bibinfo{number}{6} (\bibinfo{year}{1999}),
  \bibinfo{pages}{233--242}.
\newblock


\bibitem[\protect\citeauthoryear{Shafik, Yang, Das, Maeda-Nunez, Merrett, and
  Al-Hashimi}{Shafik et~al\mbox{.}}{2015}]%
        {shafik2015learning}
\bibfield{author}{\bibinfo{person}{Rishad~A Shafik}, \bibinfo{person}{Sheng
  Yang}, \bibinfo{person}{Anup Das}, \bibinfo{person}{Luis~A Maeda-Nunez},
  \bibinfo{person}{Geoff~V Merrett}, {and} \bibinfo{person}{Bashir~M
  Al-Hashimi}.} \bibinfo{year}{2015}\natexlab{}.
\newblock \showarticletitle{{Learning transfer-based adaptive energy
  minimization in embedded systems}}.
\newblock \bibinfo{journal}{\emph{IEEE Transactions on Computer-Aided Design of
  Integrated Circuits and Systems}} \bibinfo{volume}{35}, \bibinfo{number}{6}
  (\bibinfo{year}{2015}), \bibinfo{pages}{877--890}.
\newblock


\bibitem[\protect\citeauthoryear{Singh, Basireddy, Prakash, Merrett, and
  Al-Hashimi}{Singh et~al\mbox{.}}{2019}]%
        {singh2019collaborative}
\bibfield{author}{\bibinfo{person}{Amit Singh},
  \bibinfo{person}{Karunakar~Reddy Basireddy}, \bibinfo{person}{Alok Prakash},
  \bibinfo{person}{Geoff Merrett}, {and} \bibinfo{person}{Bashir~M
  Al-Hashimi}.} \bibinfo{year}{2019}\natexlab{}.
\newblock \showarticletitle{{Collaborative Adaptation for Energy-Efficient
  Heterogeneous Mobile SoCs}}.
\newblock \bibinfo{journal}{\emph{IEEE Trans. Comput.}} (\bibinfo{year}{2019}).
\newblock


\bibitem[\protect\citeauthoryear{Statista}{Statista}{[n. d.]}]%
        {Statista2018_apps}
\bibfield{author}{\bibinfo{person}{Statista}.} \bibinfo{year}{[n.
  d.]}\natexlab{}.
\newblock \bibinfo{title}{{Mobile App Usage - Statistics \& Facts}}.
\newblock
\newblock
\newblock
\shownote{\url{https://www.statista.com/topics/1002/mobile-app-usage/} Accessed
  24 Nov. 2018.}


\bibitem[\protect\citeauthoryear{Sultan, Ananthanarayanan, and Sarangi}{Sultan
  et~al\mbox{.}}{2014}]%
        {sultan2014processor}
\bibfield{author}{\bibinfo{person}{Hameedah Sultan}, \bibinfo{person}{Gayathri
  Ananthanarayanan}, {and} \bibinfo{person}{Smruti~R Sarangi}.}
  \bibinfo{year}{2014}\natexlab{}.
\newblock \showarticletitle{{Processor Power Estimation Techniques: A Survey.}}
\newblock \bibinfo{journal}{\emph{IJHPSA}} \bibinfo{volume}{5},
  \bibinfo{number}{2} (\bibinfo{year}{2014}), \bibinfo{pages}{93--114}.
\newblock


\bibitem[\protect\citeauthoryear{Sun, Venkatraman, Gordon, Boots, and
  Bagnell}{Sun et~al\mbox{.}}{2017}]%
        {sun2017deeply}
\bibfield{author}{\bibinfo{person}{Wen Sun}, \bibinfo{person}{Arun
  Venkatraman}, \bibinfo{person}{Geoffrey~J Gordon}, \bibinfo{person}{Byron
  Boots}, {and} \bibinfo{person}{J~Andrew Bagnell}.}
  \bibinfo{year}{2017}\natexlab{}.
\newblock \showarticletitle{{Deeply {A}ggre{V}a{T}e{D}: Differentiable
  Imitation Learning for Sequential Prediction}}. In
  \bibinfo{booktitle}{\emph{Proc. 34th Int. Conf. Machine Learning}},
  Vol.~\bibinfo{volume}{70}. \bibinfo{pages}{3309--3318}.
\newblock


\bibitem[\protect\citeauthoryear{Sutton and Barto}{Sutton and Barto}{2018}]%
        {sutton2018reinforcement}
\bibfield{author}{\bibinfo{person}{Richard~S Sutton} {and}
  \bibinfo{person}{Andrew~G Barto}.} \bibinfo{year}{2018}\natexlab{}.
\newblock \bibinfo{booktitle}{\emph{{Reinforcement Learning: An
  Introduction}}}.
\newblock \bibinfo{publisher}{MIT press}.
\newblock


\bibitem[\protect\citeauthoryear{Thomas et~al\mbox{.}}{Thomas
  et~al\mbox{.}}{2014}]%
        {thomas2014cortexsuite}
\bibfield{author}{\bibinfo{person}{Shelby Thomas} {et~al\mbox{.}}}
  \bibinfo{year}{2014}\natexlab{}.
\newblock \showarticletitle{{CortexSuite: A Synthetic Brain Benchmark Suite.}}.
  In \bibinfo{booktitle}{\emph{IISWC}}. \bibinfo{pages}{76--79}.
\newblock


\bibitem[\protect\citeauthoryear{Tian, Wang, Li, Yang, Maeda, and Xu}{Tian
  et~al\mbox{.}}{2018}]%
        {tian2018multi}
\bibfield{author}{\bibinfo{person}{Zhongyuan Tian}, \bibinfo{person}{Zhe Wang},
  \bibinfo{person}{Haoran Li}, \bibinfo{person}{Peng Yang},
  \bibinfo{person}{Rafael Kioji~Vivas Maeda}, {and} \bibinfo{person}{Jiang
  Xu}.} \bibinfo{year}{2018}\natexlab{}.
\newblock \showarticletitle{{Multi-device collaborative management through
  knowledge sharing}}. In \bibinfo{booktitle}{\emph{2018 23rd Asia and South
  Pacific Design Automation Conference (ASP-DAC)}}. \bibinfo{pages}{22--27}.
\newblock


\bibitem[\protect\citeauthoryear{ul~Islam and Lin}{ul~Islam and Lin}{2017}]%
        {ul2017hybrid}
\bibfield{author}{\bibinfo{person}{F.M.M. ul Islam} {and} \bibinfo{person}{Man
  Lin}.} \bibinfo{year}{2017}\natexlab{}.
\newblock \showarticletitle{{Hybrid DVFS Scheduling For Real-Time Systems Based
  On Reinforcement Learning}}.
\newblock \bibinfo{journal}{\emph{IEEE Systems J.}} \bibinfo{volume}{11},
  \bibinfo{number}{2} (\bibinfo{year}{2017}), \bibinfo{pages}{931--940}.
\newblock


\bibitem[\protect\citeauthoryear{Vallina-Rodriguez and
  Crowcroft}{Vallina-Rodriguez and Crowcroft}{2012}]%
        {vallina2012energy}
\bibfield{author}{\bibinfo{person}{N. Vallina-Rodriguez} {and}
  \bibinfo{person}{Jon Crowcroft}.} \bibinfo{year}{2012}\natexlab{}.
\newblock \showarticletitle{{Energy Management Techniques in Modern Mobile
  Handsets}}.
\newblock \bibinfo{journal}{\emph{IEEE Comm. Surveys \& Tutorials}}
  \bibinfo{volume}{15}, \bibinfo{number}{1} (\bibinfo{year}{2012}),
  \bibinfo{pages}{1--20}.
\newblock


\bibitem[\protect\citeauthoryear{Won et~al\mbox{.}}{Won et~al\mbox{.}}{2014}]%
        {won2014up}
\bibfield{author}{\bibinfo{person}{Jae-Yeon Won} {et~al\mbox{.}}}
  \bibinfo{year}{2014}\natexlab{}.
\newblock \showarticletitle{{Up by their bootstraps: Online learning in
  artificial neural networks for CMP uncore power management}}. In
  \bibinfo{booktitle}{\emph{2014 IEEE 20th Intl. Symp. on HPCA}}.
  \bibinfo{pages}{308--319}.
\newblock


\bibitem[\protect\citeauthoryear{Ye, Vijaykrishnan, Kandemir, and Irwin}{Ye
  et~al\mbox{.}}{2000}]%
        {ye2000design}
\bibfield{author}{\bibinfo{person}{Wu Ye}, \bibinfo{person}{Narayanan
  Vijaykrishnan}, \bibinfo{person}{Mahmut Kandemir}, {and}
  \bibinfo{person}{Mary~Jane Irwin}.} \bibinfo{year}{2000}\natexlab{}.
\newblock \showarticletitle{{The Design and Use of Simplepower: A
  Cycle-Accurate Energy Estimation Tool}}. In
  \bibinfo{booktitle}{\emph{Proceedings of the 37th Annual Design Automation
  Conference}}. ACM, \bibinfo{pages}{340--345}.
\newblock


\bibitem[\protect\citeauthoryear{Zhang et~al\mbox{.}}{Zhang
  et~al\mbox{.}}{2017}]%
        {zhang2017energy}
\bibfield{author}{\bibinfo{person}{Qingchen Zhang} {et~al\mbox{.}}}
  \bibinfo{year}{2017}\natexlab{}.
\newblock \showarticletitle{{Energy-efficient Scheduling for Real-time Systems
  Based on Deep Q-learning Model}}.
\newblock \bibinfo{journal}{\emph{IEEE Trans. on Sustainable Computing}}
  (\bibinfo{year}{2017}).
\newblock


\bibitem[\protect\citeauthoryear{Zhang, Lin, Yang, Chen, Khan, and Li}{Zhang
  et~al\mbox{.}}{2018}]%
        {zhang2018double}
\bibfield{author}{\bibinfo{person}{Qingchen Zhang}, \bibinfo{person}{Man Lin},
  \bibinfo{person}{Laurence~T Yang}, \bibinfo{person}{Zhikui Chen},
  \bibinfo{person}{Samee~U Khan}, {and} \bibinfo{person}{Peng Li}.}
  \bibinfo{year}{2018}\natexlab{}.
\newblock \showarticletitle{{A double deep Q-learning model for
  energy-efficient edge scheduling}}.
\newblock \bibinfo{journal}{\emph{IEEE Transactions on Services Computing}}
  (\bibinfo{year}{2018}).
\newblock


\end{thebibliography}

\end{document}